\documentclass[%
 reprint,
 superscriptaddress,
 amsmath,
 amssymb,
 aps,
 pra,
]{revtex4-2}

\usepackage{graphicx}
\usepackage{dcolumn}
\usepackage{bm}

\usepackage{float}
\usepackage{blindtext}
\usepackage{amsmath,amsfonts,amsthm,physics}
\usepackage{tikz}                               
\usetikzlibrary{positioning}                    
\usetikzlibrary {shapes.geometric}              
\usetikzlibrary { decorations.pathmorphing, decorations.pathreplacing, decorations.shapes, }
\graphicspath{ {./images/} }                    
\usepackage{comment}                            
\usepackage{hyperref}

\newcommand{\bs}[1]{\boldsymbol{#1}}
\newcommand{\LL}{\left}
\newcommand{\RR}{\right}
\newcommand{\id}{\mathbb{I}}
\newcommand{\sz}{\sigma_Z}
\newcommand{\de}{\text{d}}
\newcommand{\e}{\varepsilon}

\renewcommand{\selectlanguage}[1]{}

\begin{document}

\preprint{APS/123-QED}

\title{Enhanced simultaneous quantum-classical communications under composable security}

\author{Nicholas Zaunders} \email{n.zaunders@uq.edu.au}
\affiliation{
    Centre for Quantum Computation and Communication Technology, School of Mathematics and Physics, University of Queensland, St Lucia, Queensland 4072, Australia.
}
\affiliation{
    School of Electrical Engineering and Telecommunications, University of New South Wales, Sydney, NSW 2052, Australia.
}

\author{Ziqing Wang}
\affiliation{
    School of Electrical Engineering and Telecommunications, University of New South Wales, Sydney, NSW 2052, Australia.
}

\author{Robert Malaney}
\affiliation{
    School of Electrical Engineering and Telecommunications, University of New South Wales, Sydney, NSW 2052, Australia.
}

\author{Ryan Aguinaldo}
\affiliation{
    Northrop Grumman Corporation, San Diego, CA 92128, USA.
}

\author{Timothy C. Ralph}
\affiliation{
    Centre for Quantum Computation and Communication Technology, School of Mathematics and Physics, University of Queensland, St Lucia, Queensland 4072, Australia.
}

\begin{abstract}
Simultaneous quantum-classical communications (SQCC) protocols are a family of continuous-variable quantum key distribution (CV-QKD) protocols which allow for quantum and classical symbols to be integrated concurrently on the same optical pulse and mode. In this work, we present a revised analysis of simultaneous quantum-classical communications in Gaussian-modulated coherent-state CV-QKD protocols. We address security concerns inherently associated with SQCC schemes and provide an updated model of the coupling between the classical and quantum channels. We provide evidence for our model via Monte Carlo simulation. We compute the performance of our revised SQCC protocol in terms of the secret-key generation rate optimised over free parameters and demonstrate improved quantum efficiency for a given classical bit-error rate. Lastly, we extend our analysis into the finite-key regime, where we propose a scheme for composably-secure SQCC under realistic operating conditions and demonstrate that our scheme retains the advantage in quantum performance over previous models.
\end{abstract}

\maketitle

\section{Introduction} \label{sec:intro}
Quantum key distribution (QKD) is a contemporary quantum technology that exploits quantum uncertainty to offer unconditionally information-theoretic secure key exchange between two parties, regardless of an eavesdropper's resources \cite{bennett_quantum_2014, shor_simple_2000}. In QKD protocols, random one-time-pad \cite{shannon_communication_1949} bits are encoded into nonorthogonal quantum states \cite{bennett_quantum_1992}, which cannot be measured by a malicious attacker without introducing a detectable error onto the signal. While QKD can in principle be performed with a variety of different basis states, the choice of basis in which secret-key bits are encoded can have a significant impact on the effectiveness of a specific protocol, and a number of different protocols have been proposed with the aim of exploiting the advantages and minimizing the disadvantages associated with any particular scheme. Examples include the discrete-variable (DV-QKD) protocols \cite{pirandola_advances_2020}, such as BB84 \cite{bennett_quantum_2014} and the Lo-Chau protocol \cite{lo_unconditional_1999} as well as more modern protocols such as the twin-field protocol \cite{lucamarini_overcoming_2018}, which encode quantum data into finite-dimensional Hilbert spaces such as photon polarisation.

Alternatively, continuous-variable (CV-QKD) protocols encode secret bits in the basis of some infinite-dimensional Hilbert space, most commonly the continuous quadrature phase space of the quantized electromagnetic field \cite{ralph_continuous_1999}. Of the continuous-variable protocols, the most ubiquitous are those which encode quantum information via a Gaussian ensemble of coherent states \cite{grosshans_continuous_2002}. CV protocols offer a practical advantage over DV protocols: coherent state generation is efficient, fast, simple and robust, with stable states able to be generated at up to 100 MHz repetition rates in practical protocols \cite{zhang_long-distance_2020, kumar_experimental_2019, jain_practical_2022, tang_measurement-device-independent_2014} using existing off-the-shelf telecommunications components and without requiring cryogenic temperatures. Similarly, state measurement in these protocols is performed via homodyne or heterodyne detection, which can also be done efficiently at high repetition rates using off-the-shelf components without the need for cryogenics. 

A second advantage of Gaussian coherent-state QKD protocols is their capacity for integration with classical coherent-state communications protocols, termed simultaneous quantum-classical communication (SQCC) \cite{qi_simultaneous_2016}. SQCC operates by encoding classical data via the mean value of the Gaussian distribution encoding the quantum symbols, where the small random phase-space fluctuation containing the quantum information is appended onto the much larger pre-determined phase-space displacement representing the classical symbol. This combined scheme is unique to CV-QKD and offers an invaluable advantage to any practical deployment of a QKD network using continuous variables, as SQCC effectively increases the number of available channels in any network without incurring any additional overhead. This advantage is especially salient when considering the possibility of freely appending quantum key capabilities onto existing classical communications networks, or performing QKD on satellite platforms where constraints on physical size and energy consumption motivate using the same channel and optical mode to perform both the quantum and classical signalling \cite{zaunders_quantum-amplified_2024}.

Despite these advantages, the current body of work on SQCC protocols shows much potential for further development. Previous proposals \cite{qi_simultaneous_2016, qi_noise_2018, pan_simultaneous_2020, zaunders_quantum-amplified_2024, winnel_classical-quantum_2024} consider a simplified model of the quantum-classical channel coupling, sufficient to capture protocol performance only at a high classical quality-of-service, and do not explicitly present a full security analysis. Furthermore, all contemporary works only present results for the simplified case of a shared key of infinite size, which is not a realistic model of practical implementations of QKD protocols \cite{pirandola_advances_2020}.

In this work, we propose an improved model for SQCC CV-QKD protocols under the requirement of composable security in the finite-size regime, and show that our scheme presents an improved security analysis relative to the previous leading SQCC model. We then proceed to provide an analytic description of the manner in which the errors associated with the classical measurement affects both the security requirements and numerical behaviour of the combined protocol. We demonstrate that the net effect of classical bit errors is to reduce the variance of the quantum measurement distribution in a non-Gaussian manner, in contradiction to previous models suggesting a Gaussian broadening, and show that directly implementing SQCC protocols without proper consideration of the output state can produce non-physical statistics, potentially rendering the protocol insecure. We confirm the predictions of our analytic model via Monte Carlo simulation, and subsequently close the security loophole posed by the non-physical data by physicalising the SQCC process via the introduction of an additional renormalisation step. We use our now-secure model to numerically derive the optimal secret key generation rate in the asymptotic regime and demonstrate a clear advantage in the quantum performance for a given classical quality-of-service when compared to the previous scheme. We also present equivalent results for the finite-key regime and demonstrate the security of the improved model under realistic conditions. The advantage of the model over previous works is also maintained in the finite-key regime.

The remainder of this paper is structured as follows. In Section \ref{sec:protocol_description}, we briefly summarise the general operation of SQCC protocols and propose the specific scheme which we intend to analyse. We also address specific security concerns and propose appropriate solutions. In Section \ref{sec:results}, we present firstly the analytic results associated with our model, detailing the explicit effect of the classical signal on the quantum data and the need and implementation of the subsequent renormalisation (Sections \ref{sec:results_analytic_cv}, \ref{sec:results_normalisation}), and then provide evidence for the analytic model via simulation (Section \ref{sec:results_montecarlo}). We then present the optimized key rate of the protocol in both the asymptotic (Section \ref{sec:results_asymptotic}) and finite-size regimes (Section \ref{sec:results_finitesize}). We present some concluding remarks in Section \ref{sec:discussion}. Further mathematical detail and additional arguments relevant to the main body are included in Appendices \ref{app:qfunction}-\ref{app:fnsz_keyrate}.

\section{Protocol description and security analysis} \label{sec:protocol_description}
In this section, we will first briefly illustrate the general operating procedure of SQCC protocols in the continuous-variable space and describe the method by which the classical and quantum protocols are conjoined, measured, and processed at the receiver. In the second half, we will characterise the specific structure of the protocol used in this work in greater detail, with a particular emphasis on the rectification of security concerns present in other protocols utilising SQCC.

\subsection{General formulation of SQCC protocols in the prepare-and-measure basis} \label{sec:protocol_description_general}
In keeping with previous works \cite{qi_simultaneous_2016, qi_noise_2018, pan_simultaneous_2020, winnel_classical-quantum_2024}, we elect to model our SQCC system as a general combination of a Gaussian-modulated coherent-state (GMCS) CV-QKD protocol with heterodyne detection \cite{weedbrook_quantum_2004}, for transmitting quantum information, and some arbitrary classical coherent-state communications protocol for transmitting classical binary data. The protocol in the prepare-and-measure basis proceeds as follows:
\begin{enumerate}
    \item Alice begins by randomly drawing some value $\tilde \alpha$ from a zero-mean bivariate Gaussian distribution of variance $V$, and prepares the coherent state $\ket{\tilde \alpha}$. This represents the quantum symbol used for key generation.
    \item She then displaces her state by some large phase-space displacement $\tilde d$, which she draws from her classical alphabet $\{ \tilde d_i \}$ and associates with some classical symbol.
    \item The state $|\tilde \alpha + \tilde d \rangle$, containing the joint quantum-classical symbol $\tilde \alpha + \tilde d$, is then sent over a lossy bosonic channel $\mathcal{E}(T, \e)$ with transmissivity $T$ and thermal-noise input $\e$.
    \item Bob performs heterodyne detection on the received state and obtains some complex measurement outcome $\tilde \beta$. This outcome is a Gaussian random variable with mean value centered on $\sqrt{T}d$ and is correlated with Alice's prepared quantum symbol $\tilde \alpha$.
    \item Bob now attempts to recover the two data streams sent to him by Alice. He does this by separating his measurement outcomes into two parts: the small Gaussian fluctuations associated with the quantum information $\tilde \alpha$ and the large displacement $\tilde d$ associated with the classical information. The easiest way for Bob to do this is to first attempt to identify the large classical displacement of any shot via a threshold discrimination, i.e. discriminating between classical symbols based on what region of the phase space they are measured to be in. Given some knowledge of the classical protocol, Bob can then erase the effect of that large displacement by applying the inverse displacement to the data point (re-displacement). This postprocessing procedure allows Bob to reproduce the classical data stream as well as emulate the measurement outcomes that would have been generated had Alice performed ordinary zero-mean GMCS CV-QKD instead.
    \item If necessary, Bob renormalises his data by applying an electronic gain, scaling his measurement results by some constant factor. This process allows Bob to identify the Gaussian-order characteristics of his non-Gaussian SQCC protocol with an equivalent virtual Gaussian protocol, from which he infers the mutual information and Holevo quantity.
    \item Bob proceeds with the remaining steps (information reconciliation, parameter estimation, privacy amplification, etc.) required to distill a secret key out of the new set of renormalised emulated zero-mean quantum measurements $\{ \tilde \beta' \}$ under reverse reconciliation.
\end{enumerate}

\subsection{Security analysis}
\label{sec:protocol_description_specific}
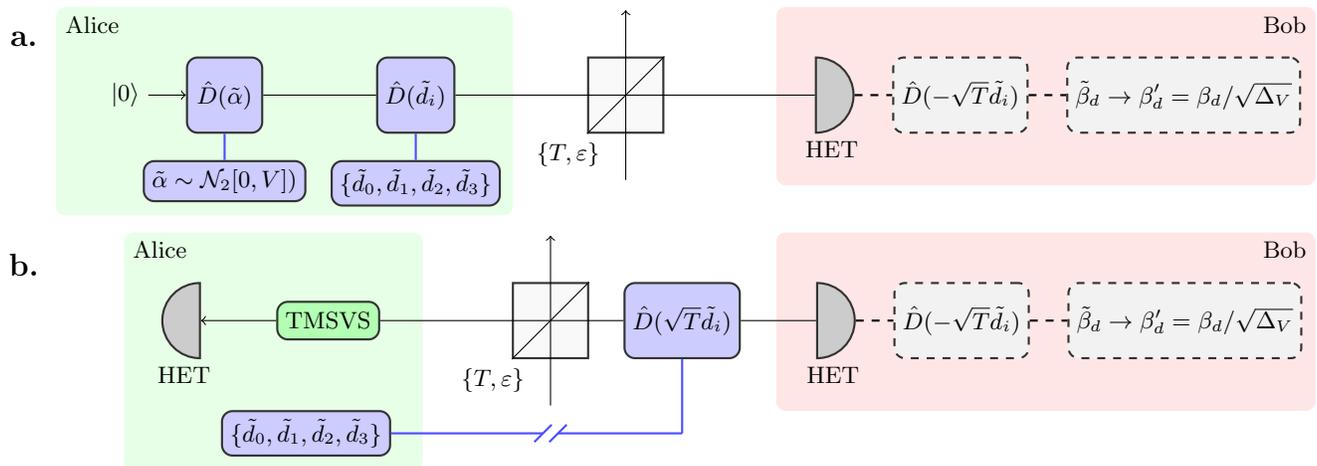
\begin{figure*}
    \centering
    \begin{tikzpicture}[
    RDET/.style={
    semicircle, draw=black!80, fill=gray!40, thick, minimum size=5mm, rotate = 270
    },
    LDET/.style={
    semicircle, draw=black!80, fill=gray!40, thick, minimum size=5mm, rotate = 90
    },
    TMSVS/.style={
    rectangle, rounded corners, draw=black!80, fill=green!30, thick, minimum size=5mm
    },
    CHANNEL/.style={
    rectangle, draw=black!80, fill=gray!5, thick, minimum size=10mm
    },
    BOBCOVER/.style={
    rectangle, rounded corners, draw=white!0, fill=red!10, thick, inner xsep=36mm, inner ysep=12mm
    },
    ALICECOVER/.style={
    rectangle, rounded corners, draw=white!80, fill=green!10, thick, inner xsep=30.5mm, inner ysep=14mm
    },
    ALICECOVER2/.style={
    rectangle, rounded corners, draw=white!80, fill=green!10, thick, inner xsep=20mm, inner ysep=16mm
    },
    DISPOP/.style={
    rectangle, rounded corners, draw=black!80, fill=blue!20, thick, minimum size=10mm
    },
    POSTDISPOP/.style={
    rectangle, rounded corners, draw=black!80, dashed, fill=gray!10, thick, minimum size=10mm
    },
    CLASSICALENCODE/.style={
    rectangle, rounded corners, draw=black!80, fill=blue!20, thick, minimum size=5mm
    },
    LABEL/.style={
    rectangle
    }
    ]

    \node[LABEL](eb) at (-7.0, 0.75){\large{\textbf{a.}}};
    
    \node[ALICECOVER](aliceCover)
    at (-6.6, 1.2) [anchor = north west] 
    {};
    \node[LABEL](AliceL) 
    [below right = 0.05cm of aliceCover.north west] [anchor=north west]
    {Alice};

    \node[BOBCOVER](bobCover)
    at (10.2, 1.2) [anchor = north east] 
    {};
    \node[LABEL](AliceL) 
    [below left = 0.05cm of bobCover.north east] [anchor=north east] 
    {Bob};

    \node[CHANNEL](channel)
    at (1, 0) [anchor = center] 
    {};
    \node[LABEL](channelCharacteristics)
    [left = 0.25cm of channel.south west] [anchor=north] 
    {$\{ T, \e \}$};
    \node[LABEL](channelIn)
    [below = 0.5cm of channel]
    {};
    \node[LABEL](channelOut)
    [above = 0.5cm of channel]
    {};
    
    \node[DISPOP](displace) 
    [left = 1.75cm of channel]
    {$\hat D(\tilde d_i)$};
    \node[CLASSICALENCODE](classicalEncoder) 
    [below = 0.35cm of displace]
    {$\{\tilde d_{0}, \tilde d_{1}, \tilde d_{2}, \tilde d_{3}\}$};
    \node[DISPOP](TMSVS) 
    [left = 1.5cm of displace]
    {$\hat D(\tilde \alpha)$};
    \node[CLASSICALENCODE](quantumEncoder) 
    [below = 0.35cm of TMSVS]
    {$\tilde \alpha \sim \mathcal{N}_2[0, V])$};
    \node[LABEL](aliceVacuumInputLabel) 
    [left = 0.5cm of TMSVS]
    {$\ket{0}$};

    \node[RDET](bobHet)
    [right = 2cm of channel] [anchor=chord center]
    {};
    \node[LABEL](bobHetL)
    [below = 0.5cm of bobHet.center] [anchor=north]
    {HET};
    \node[POSTDISPOP](redisplacePost) 
    [right = 0.5cm of bobHet.north]
    {$\hat D(-\sqrt{T} \tilde d_i)$};
    \node[POSTDISPOP](renormalisePost) 
    [right = 0.5cm of redisplacePost]
    {$\tilde \beta_d \rightarrow \beta_d' = \beta_d/\sqrt{\Delta_V}$};
    
    \draw[->] (aliceVacuumInputLabel.east) -- (TMSVS.west);
    \draw[-] (TMSVS.east) -- (displace.west);
    \draw[-] (displace.east) -- (bobHet.chord center);
    \draw[->] (channelIn.center) -- (channelOut.center);
    \draw[-, thick, dashed] (bobHet.north) -- (redisplacePost.west);
    \draw[-, thick, dashed] (redisplacePost.east) -- (renormalisePost.west);
    
    \draw[-, blue!70, thick] (quantumEncoder.north) -- (TMSVS.south);
    \draw[-, blue!70, thick] (classicalEncoder.north) -- (displace.south);
    
    \draw[-] (channel.south west) -- (channel.north east);

    \node[LABEL](eb) at (-7.0, -2.25){\large{\textbf{b.}}};
    
    \node[ALICECOVER2](aliceCover)
    at (-5.7, -1.8) [anchor = north west] 
    {};
    \node[LABEL](AliceL) 
    [below right = 0.05cm of aliceCover.north west] [anchor=north west]
    {Alice};

    \node[BOBCOVER](bobCover)
    at (10.2, -1.8) [anchor = north east] 
    {};
    \node[LABEL](AliceL) 
    [below left = 0.05cm of bobCover.north east] [anchor=north east] 
    {Bob};

    \node[CHANNEL](channel)
    at (0, -3) [anchor = center] 
    {};
    \node[LABEL](channelCharacteristics)
    [left = 0.25cm of channel.south west] [anchor=north] 
    {$\{ T, \e \}$};
    \node[LABEL](channelIn)
    [below = 0.5cm of channel]
    {};
    \node[LABEL](channelOut)
    [above = 0.5cm of channel]
    {};
    
    \node[TMSVS](TMSVS) 
    [left = 1.75cm of channel]
    {TMSVS};
    \node[CLASSICALENCODE](classicalEncoder) 
    at (-3.25, -4.5) [anchor = center]
    {$\{\tilde d_{0}, \tilde d_{1}, \tilde d_{2}, \tilde d_{3}\}$};
    \node[LDET](aliceHet) 
    [left = 1.0cm of TMSVS] [anchor=chord center] 
    {};
    \node[LABEL](aliceHetL)
    [below = 0.5cm of aliceHet.center] [anchor=north]
    {HET};

    \node[DISPOP](displace) 
    at (1.75, -3) [anchor = center]
    {$\hat D(\sqrt{T} \tilde d_i)$};
    \node[RDET](bobHet)
    [right = 1.0cm of displace] [anchor=chord center]
    {};
    \node[LABEL](bobHetL)
    [below = 0.5cm of bobHet.center] [anchor=north]
    {HET};
    \node[POSTDISPOP](redisplacePost) 
    [right = 0.5cm of bobHet.north]
    {$\hat D(-\sqrt{T} \tilde d_i)$};
    \node[POSTDISPOP](renormalisePost) 
    [right = 0.5cm of redisplacePost]
    {$\tilde \beta_d \rightarrow \beta_d' = \beta_d/\sqrt{\Delta_V}$};

    \node[LABEL](cChannel1)
    at (-0.1, -4.5)
    {};
    \node[LABEL](cChannel2)
    at (0.1, -4.5)
    {};
    \node[LABEL](cChannel3)
    at (1.75, -4.5)
    {};
    
    \draw[->] (TMSVS.west) -- (aliceHet.chord center);
    \draw[-] (TMSVS.east) -- (displace.west);
    \draw[-] (displace.east) -- (bobHet.chord center);
    \draw[->] (channelIn.center) -- (channelOut.center);
    \draw[-, thick, dashed] (bobHet.north) -- (redisplacePost.west);
    \draw[-, thick, dashed] (redisplacePost.east) -- (renormalisePost.west);
    
    \draw[-, blue!70, thick] (classicalEncoder.east) -- (cChannel1.center);
    \draw[-, blue!70, thick] (cChannel2.center) -- (cChannel3.center);
    \draw[-, blue!70, thick] (cChannel3.center) -- (displace.south);
    \draw[-, blue!70, thick] (cChannel1.south west) -- (cChannel1.north east);
    \draw[-, blue!70, thick] (cChannel2.south west) -- (cChannel2.north east);
    
    \draw[-] (channel.south west) -- (channel.north east);
    
\end{tikzpicture}
    \caption{Protocol diagram describing the secure SQCC scheme described in Section \ref{sec:protocol_description} in the prepare-and-measure picture (\textbf{a}), and the equivalent entanglement-based picture (\textbf{b}) used for security analysis. In the entanglement-based picture, Alice shares one mode of a two-mode squeezed vacuum state (TMSVS), which she passes through a noisy channel $\mathcal{E}(T,\e)$ before displacing it by some large displacement to encode a classical symbol. We model Alice's classical displacement as occurring after the outgoing TMSVS mode is passed through the channel to encode the assumption that Eve has perfect knowledge of the classical communications, and so can reconstruct Alice's original Gaussian state from the combined signal with perfect fidelity. This allows Alice and Bob to infer Eve's optimal attack, and consequently her information, via Gaussian optimality. Bob then measures the outgoing mode, emulates Alice's original quantum signal via threshold discrimination and re-displacement, and renormalises the data to produce a measurement distribution consistent with a Gaussian channel, from which Alice and Bob infer Eve's information and distill a shared secret key.}
    \label{fig:protocol_diagram}
\end{figure*}

In order to evaluate the security of the prepare-and-measure SQCC protocol described in Section \ref{sec:protocol_description_general} (Fig.~\ref{fig:protocol_diagram}[a]), we must first convert it into the equivalent virtual entanglement-based scheme (Fig.~\ref{fig:protocol_diagram}[b]). Alice begins by generating a two-mode squeezed vacuum state of squeezing $0 \leq \lambda \leq 1$, which is a zero-mean Gaussian state with covariance matrix \cite{weedbrook_gaussian_2012}
\begin{align}
    \bs{V} &= \begin{pmatrix}
        V\id & \sqrt{V^2 - 1}\sz \\
        \sqrt{V^2 - 1}\sz & V\id
    \end{pmatrix},
\end{align}
for $V = (1+\lambda^2)/(1 - \lambda^2)$ and Pauli matrix $\sz$. Alice then retains one mode and displaces the other according to the procedure described in Section \ref{sec:protocol_description_general}. The choice of classical alphabet by Alice is not a trivial one: in order to guarantee composable security in the finite-key regime, we are required to maintain symmetry between the quadrature operators $\hat q$ and $\hat p$ \cite{leverrier_composable_2015}. This therefore rules out binary protocols, such as binary phase-shift-keying, which are manifestly asymmetric. For ease, we choose the simplest symmetric classical protocol to encode the classical data, which in this instance is quadrature phase-shift-keying (QPSK). QPSK has the alphabet
\begin{align}
    \tilde d \in \{d_k = d e^{\frac{i\pi (2k - 1)}{4}} : d \geq 0, k = 1,2,3,4 \}, \label{eq:classical_alphabet}
\end{align}
corresponding to the binary symbols $00$, $01$, $11$ and $10$ respectively for $k = 1,2,3,4$. QPSK may also be visualised as assigning a classical symbol to each quadrant of the complex plane, where the first quadrant is identified with $00$, the second quadrant to $01$, and so on.

After the outgoing mode is displaced according to the desired classical symbol $\tilde d$, Alice passes it through the untrusted communication channel $\mathcal{E}(T, \e)$, where we assume $\mathcal{E}$ to be a lossy bosonic channel with transmissivity $T$ and thermal-state input with mean photon number $\overline n = T\epsilon / 2(1-T)$. We also assume a reconciliation efficiency $\beta = 0.95$, which represents a reasonable cross-section of Gaussian error-correction efficiencies \cite{jain_practical_2022}.

After the channel, Alice and Bob share a mixed two-mode entangled Gaussian state fully described by a mean vector $\bs \mu$ and covariance matrix $\bs V$ \cite{laudenbach_continuousvariable_2018}:
\begin{align}
    \bs \mu &= \left( 0, 0, \sqrt{T}\Re\{\tilde d\}, \sqrt{T}\Im\{\tilde d\} \right)^T, \label{eq:init_mean}\\
    \bs V &= \begin{pmatrix} 
        V \id & \sqrt{T(V^2 - 1)}\sz \\
        \sqrt{T(V^2 - 1)}\sz & [T(V + \e - 1) + 1]\id 
    \end{pmatrix} \label{eq:init_covar}\\
    &\equiv \begin{pmatrix} 
        a \id & c \sz \\ 
        c \sz & b\id 
    \end{pmatrix}.
\end{align}
Alice and Bob now perform heterodyne detection on their respective modes and obtain complex measurement outcomes $\tilde \alpha \equiv \alpha_x + i\alpha_y$ and $\tilde \beta \equiv \beta_x + i\beta_y$ respectively. By construction, their measurement results are distributed according to the Husimi Q-function of the state described by the mean vector $\bs{\mu}$ and covariance matrix $\bs{V}$ given in Eqs. (\ref{eq:init_mean},~\ref{eq:init_covar}). Their joint measurement ensemble therefore has the probability density function (Appendix \ref{app:qfunction})
\begin{align}
    Q(\tilde \alpha, \tilde \beta) = Q^{\bs \mu}_{\bs V}(\alpha_x, \alpha_y, \beta_x, \beta_y).
\end{align}
At this stage, Bob emulates the equivalent zero-mean CV-QKD results $\{ \tilde \beta_d \}$ by postprocessing his measurement outcomes according to Section \ref{sec:protocol_description_general}, obtaining a non-Gaussian measurement distribution represented by the non-Gaussian Q-function $Q_d$ (Appendix \ref{app:covar}). However, in order to accurately estimate Eve's information from the joint state held by Alice and Bob via Gaussian optimality, Alice and Bob must ensure the Gaussian characteristics of their shared non-Gaussian state $Q_d$ correspond to an equivalent physically-realizable Gaussian protocol, which they can assume Eve purifies in order to infer her Holevo information \cite{navascues_optimality_2006, garcia-patron_unconditional_2006}. Bob does this by rescaling his measurement values by some factor $1/\sqrt{\Delta_V}$ in such a way that the total action of the channel, measurement, postprocessing and rescaling on the measurement distribution always produces statistics equivalent to some physically-legitimate virtual Gaussian protocol with effective channel $\mathcal{E'}$ under heterodyne measurement (Appendix \ref{app:covar_renorm}). Only after rescaling is Bob able to exploit Gaussian optimality and extremality \cite{navascues_optimality_2006, garcia-patron_unconditional_2006, wolf_extremality_2006} to place a lower bound on the achievable secret-key length distillable from his postprocessed and rescaled data $\{ \tilde \beta_d' \}$.

\textit{A priori}, it is reasonable to assume that Alice sends a random mix of symbols when encoding her classical data for any arbitrary bit string. The total distribution Bob obtains after measuring every shot would thus be a balanced statistical mixture of the four quantum states corresponding to each possible classical symbol:
\begin{align}
    Q_{in}(\bs \alpha) &= \frac{Q^{d_1} + Q^{d_2} + Q^{d_3} + Q^{d_4}}{4}.
\end{align}
We might also justify this on a shot-by-shot basis by recalling that Bob does not know which of the four symbols he is going to receive, and so the quantum state he receives is a probabilistic mix of the four encodings. (Here we use the shorthand $Q^{d_k}$ to indicate the Q-function $Q^{\bs \mu}_{\bs V}$ of the state with covariance $\bs V$ and mean $\bs \mu$ associated with the classical displacement $\tilde d = d_k$ as per Eqs. [\ref{eq:init_mean},~\ref{eq:init_covar}]).

However, such an approach would not be secure in theory. If we treat the classical displacement as originating from Alice's laboratory, then the ensemble received by Eve is necessarily non-Gaussian, making a quantification of the optimal attack considerably more difficult \cite{navascues_optimality_2006, garcia-patron_unconditional_2006}. To resolve this, we propose to analyse the security of an equivalent virtual scheme wherein Alice performs the displacement operation on the quantum signal immediately prior to measurement by Bob via an insecure classical communications channel. The effect of this change is dual: it ensures no representations are made on the security of the classical signal, and also restricts Eve's measurement ensembles to be Gaussian, which constrains her attack according to Gaussian optimality. Specifically, it encodes the assumption that Eve has perfect knowledge of Alice's classical protocol and message, and can therefore reproduce Alice's Gaussian quantum signal from the combined classical and quantum distribution with perfect fidelity. Furthermore, since by Gaussian optimality Eve maximises her information by performing a Gaussian attack on Alice's Gaussian ensemble, the scheme described in Fig. \ref{fig:protocol_diagram}b represents the best-case scenario for Eve.

We also infer the Holevo quantity of the protocol in a specialised way. While Eve receives on a shot-by-shot basis a non-displaced quantum symbol, her knowledge of the classical register nevertheless makes it easier for her to predict Bob's heterodyne measurement result, and therefore his postprocessed measurement result as well. For example, if Eve knows an intercepted quantum measurement is associated with a classical symbol of $00$, then she knows with higher confidence that the quantum symbol Bob measures will be in the first quadrant. She cannot, however, predict the precise classical errors Bob will make upon measurement. In inferring Eve's Holevo quantity, we therefore make the assumption that Eve separates her total ensemble into four sub-ensembles, with each containing the intercepted quantum data for one specific classical symbol, and performs her attack optimally on each sub-ensemble while accounting for the existence of the classical displacement. By symmetry we know that the Holevo quantity for each sub-ensemble will be the same, and so we extrapolate that the Holevo quantity for the entire ensemble is also given by the Holevo quantity for any of the sub-ensembles. This is a reasonable assumption, since by symmetry we not expect there to be any exploitable correlations between the sub-ensembles that would allow Eve to gain an advantage by attacking the whole ensemble.

\section{Results} \label{sec:results}
In this section, we will summarise the effect of the classical encoding and decoding steps on the quantum signal, and illuminate some of the chief analytical differences between our protocol and the previous model used in the literature \cite{qi_noise_2018} via Monte Carlo simulation. We will then proceed to compute the optimised quantum performance in terms of the secret-key bits per pulse for the SQCC protocol, in both the asymptotic and finite-key regime.

\subsection{Monte Carlo simulation of signal postprocessing} \label{sec:results_montecarlo}
Before proceeding with the results proper, it is instructive to obtain an intuition for the behaviour of the measurement distribution obtained by Bob over the course of the SQCC protocol, and further to motivate the need for an accurate model of SQCC. An effective way of determining this is via Monte Carlo simulation, since by construction the quantum heterodyne measurement outcomes obtained by Bob in a real protocol are physically indistinguishable from simulated data drawn from an identical distribution. Moreover, because the key elements of the protocol are done in postprocessing, the data operations required for a Monte Carlo simulation are in essence identical to that of a `real' protocol. 

\begin{figure*}
    \centering
    \includegraphics{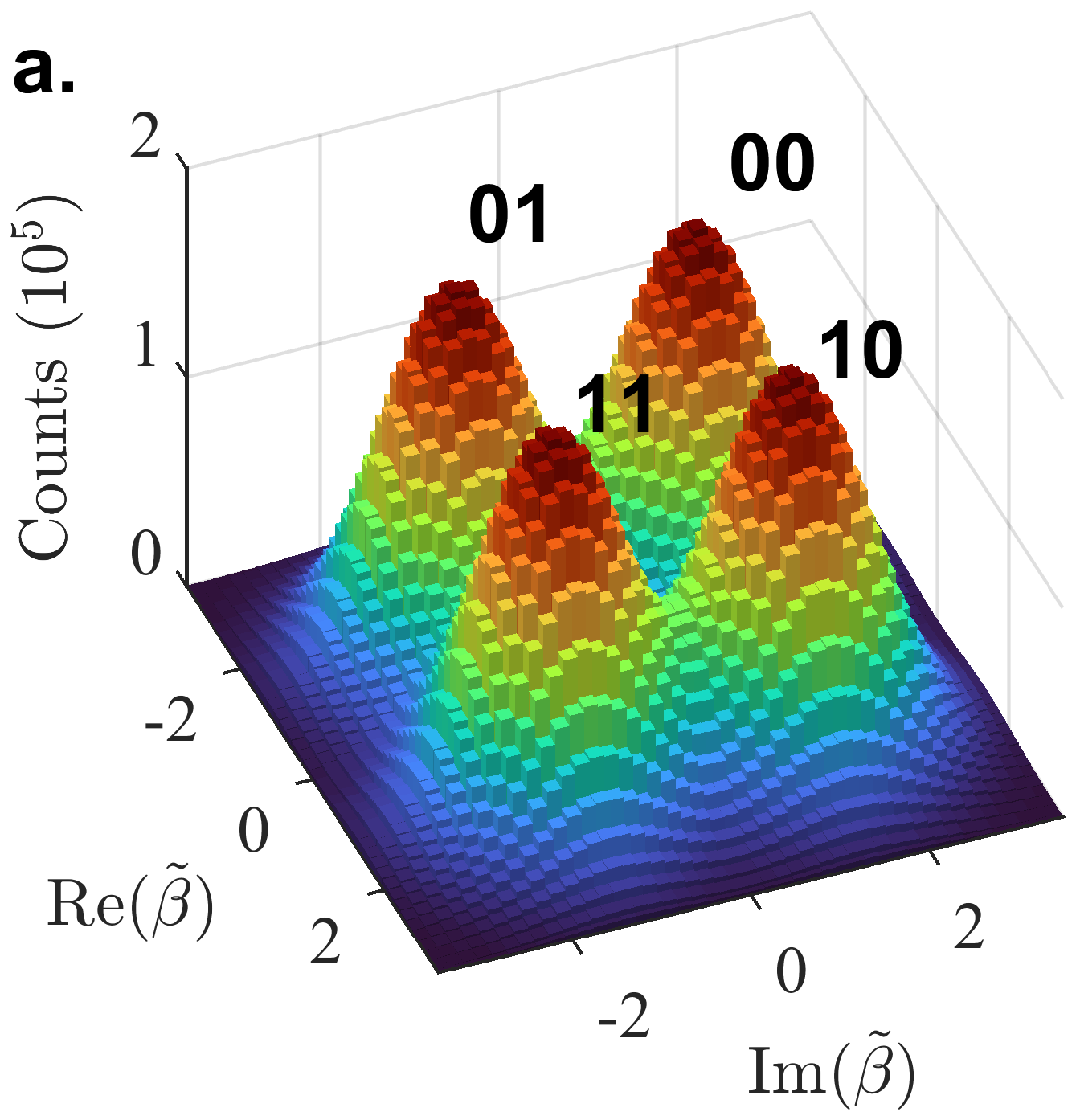}
    \includegraphics{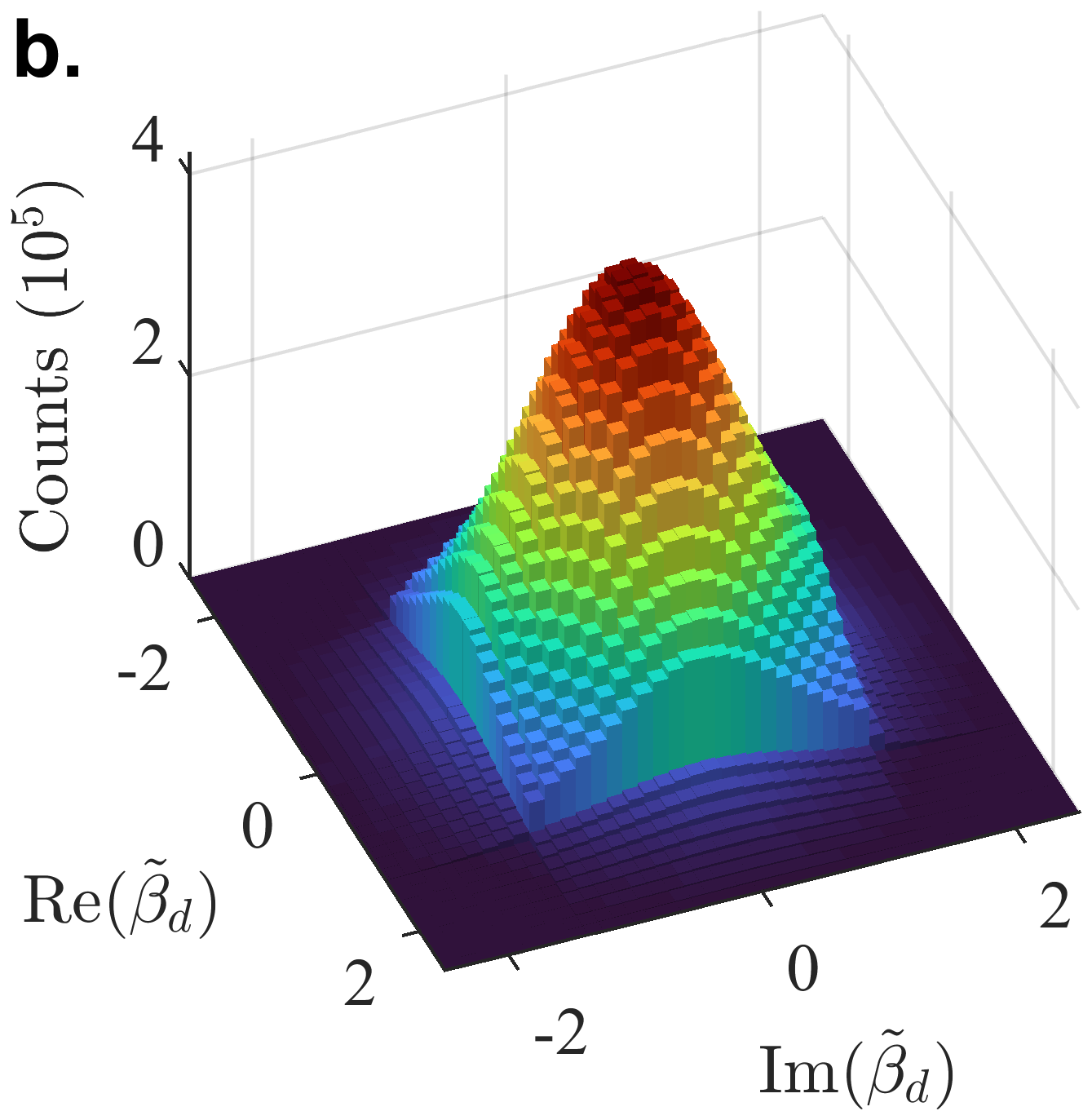}
    \includegraphics{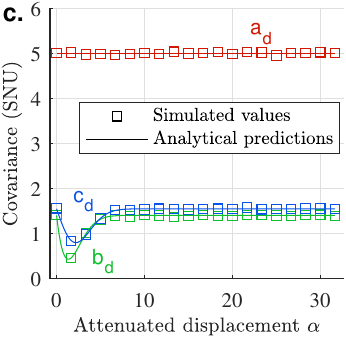}
    \caption{Monte Carlo simulations of the joint measurement distribution possessed by Alice and Bob for $N = 10^8$ shots. (\textbf{a}) A simulated distribution of Bob's data $\tilde \beta$ immediately after heterodyne measurement of the joint quantum state described by Eqs. \eqref{eq:init_mean}-\eqref{eq:init_covar}, evaluated for Alice's modulation variance $V = 5$, channel transmissivity $T = 0.1$, channel excess noise $\e = 0.05$ and initial classical displacement $d = 12$. Each combined quantum and classical symbol sent by Alice manifests as a large displacement in the phase space in one of the four cardinal directions, encoding a classical bit string, together with a smaller quantum fluctuation encoding a quantum symbol. (\textbf{b}) Identifying each symbol's classical data via the threshold discrimination process and re-displacing accordingly causes the postprocessed measurements $\tilde \beta_d$ to assume a quasi-Gaussian distribution, with the degree of non-Gaussianity proportional to the overlap of the joint quantum-classical symbols. (\textbf{c}) A comparison of the covariance matrix elements $a_d$, $b_d$ and $c_d$ characterising Alice and Bob's joint postprocessed distribution, as predicted by theory (solid lines) and as calculated empirically via generation and postprocessing of $10^5$ simulated `real' measurements (squares).}
    \label{fig:pubfig_montecarlo}
\end{figure*}

To test our model, we therefore generate a random sample of bivariate complex Gaussian data of size $N~=~10^8$ according to the mean vector and covariance matrix given in Eqs. \eqref{eq:init_mean} and \eqref{eq:init_covar} respectively, to simulate the correlated heterodyne measurement results obtained by Alice and Bob for the noisy channel $\mathcal{E}(T, \epsilon)$ (Fig. \ref{fig:pubfig_montecarlo}[a]). We postprocess the data blindly according to the threshold discrimination and re-displacement operation described in Section \ref{sec:protocol_description_general} to generate Bob's emulated zero-mean distribution (Figure \ref{fig:pubfig_montecarlo}[b]). From this emulated distribution, we directly calculate the sample bit-error rate $e_C$ and covariance matrix elements and compare them with the analytic predictions (Figure \ref{fig:pubfig_montecarlo}[c]) made in the following section.

Figures \ref{fig:pubfig_montecarlo}(b) and \ref{fig:pubfig_montecarlo}(c) usefully illuminate some general properties of SQCC protocols that lie in contradiction to assumptions made implicitly in previous works \cite{qi_noise_2018}, which describe the final matrix shared by Alice and Bob in a homodyne-measurement binary phase-shift-keyed SQCC protocol as
\begin{align}
    \bs{V} &= \begin{pmatrix}
        V\id & \sqrt{T(V^2 - 1)}\sz \\
        \sqrt{T(V^2 - 1)}\sz & \LL[ T(V + \e' - 1) + 1 \RR]\id
    \end{pmatrix},
\end{align}
where
\begin{align}
    \e' &= \e + 4\alpha^2 e_C. \label{eq:qiNoise}
\end{align}
In this formulation, the factor coupling the classical and quantum channels is the term $4\alpha^2 e_C$, and its position in the covariance matrix implies that the sole effect of bit errors on the quantum signal is equivalent to a Gaussian broadening of Bob's measurement distribution in a manner identical to an untrusted Gaussian excess-noise source. However, Figure \ref{fig:pubfig_montecarlo}(b) demonstrates that the postprocessing maps the classical mixture of Gaussian states to a distribution that is definitively non-Gaussian, a fact that has strong ramifications for the security of SQCC protocols in general and which partially motivates the alternate security proof proposed in Sec. \ref{sec:protocol_description_specific}.

Furthermore, it would be expected that bit errors on the classical signal would degrade the correlation between Alice and Bob's data as a result of the incorrect displacement operation, something which is manifest in simulated data but not accounted for in \cite{qi_noise_2018}. Similarly, we determine via simulation that the effect of the incorrect displacement actually works to reduce the variance of Bob's postprocessed distribution, rather than the broadening suggested by Eq. \eqref{eq:qiNoise}.

\subsection{Analytic form of the covariance matrix} \label{sec:results_analytic_cv}
Under the assumptions described in Section \ref{sec:protocol_description_specific}, the physical state Alice and Bob share prior to measurement is a Gaussian two-mode squeezed state with first and second moments equal to Eqs. \eqref{eq:init_mean} and \eqref{eq:init_covar} respectively:
\begin{align}
    \bs \mu &= \left( 0, 0, \frac{\alpha}{\sqrt{2}},  \frac{\alpha}{\sqrt{2}} \right)^T, \\
    \bs V &= \begin{pmatrix} 
        a \id & c \sz \\ 
        c \sz & b\id 
    \end{pmatrix}
\end{align}
where we use the shorthand $\alpha^2 \equiv T d^2$. If we define the quasi-signal-to-noise parameter 
\begin{align}
    \text{SNR} &= \frac{T d^2}{b+1} \equiv \frac{\alpha^2}{b + 1},
\end{align}
then the Gaussian-order statistics of the measurement distribution $\{\tilde \alpha, \tilde \beta_d\}$ shared by Alice and Bob after Bob performs his threshold discrimination and re-displacement operation are given by
\begin{align}
    \bs \mu_d &= \left( 0, 0, \frac{2 \alpha e_C}{\sqrt{2}},  \frac{2 \alpha e_C}{\sqrt{2}} \right)^T, \\
    \bs V_d &= \begin{pmatrix} 
        a_d \id & c_d \sz \\ 
        c_d \sz & b_d \id 
    \end{pmatrix}
\end{align}
where
\begin{align}
    a_d &= a \label{eq:dispcv_a}, \\
    b_d &= b + 2\alpha^2 e_C - 2 (b + 1)\delta - 2\alpha^2 e_C^2 \label{eq:dispcv_b}, \\
    c_d &= c(1-\delta) \label{eq:dispcv_c}
\end{align}
and
\begin{align}
    e_C &= \frac{1}{2}\text{erfc}\LL( \frac{\sqrt{\text{SNR}}}{2} \RR), \\
    \delta &= \sqrt{\frac{\text{SNR}}{\pi}} e^{-\text{SNR}/4}.
\end{align}

Here $e_C$ is also the bit-error rate of the classical signal. Figure 2(c) shows excellent agreement between simulated data and analytically predicted values, confirming the validity of the model.

\subsection{Renormalisation of the postprocessed state} \label{sec:results_normalisation}
    
It can be seen from Eq. \eqref{eq:dispcv_b} and Fig. \ref{fig:pubfig_montecarlo}(c) that if Bob were to blindly perform the re-displacement operation, he runs the risk of producing statistics which are physically illegitimate by virtue of being sub-shot-noise in both $\hat q$ and $\hat p$ simultaneously (i.e. $b < 1$). Since Alice is sending legitimate quantum states, it follows that the equivalent virtual channel which Bob infers the existence of is non-physical, rendering it essentially impossible to quantify an upper bound on Eve's information \cite{navascues_optimality_2006, garcia-patron_unconditional_2006}. Bob therefore performs a rescaling step on the postprocessed measurement results $\{\tilde \beta_d\}$
\begin{align}
    \tilde \beta_d \Longrightarrow \tilde \beta_d' &= \frac{\tilde \beta_d}{\sqrt{\Delta_V}}.
\end{align}
The first- and second-order measurement statistics of Alice and Bob's shared data $\{\tilde \alpha, \tilde \beta_d'\}$ now become
\begin{align}
    \bs \mu_{d}' &= \left( 0, 0, \frac{1}{\sqrt{\Delta_V}}\frac{2\alpha e_C}{\sqrt{2}},  \frac{1}{\sqrt{\Delta_V}}\frac{2\alpha e_C}{\sqrt{2}}\right)^T, \label{eq:dispPrimeMean}\\
    \bs V_d{'} &= \begin{pmatrix} 
        a_d \id & \frac{c_d}{\sqrt{\Delta_V}} \sz \\ 
        \frac{c_d}{\sqrt{\Delta_V}} \sz & \LL[ \frac{b_d}{\Delta_V} + \frac{1}{\Delta_V} - 1 \RR] \id
    \end{pmatrix}. \label{eq:dispPrimecovar}
\end{align}
Here Bob encounters some freedom in his choice of renormalisation. One possible option is to use the electronic amplification to restore the correlations between the two data sets, by setting his gain in such a way as to counter the correlation-reducing effect of the postprocessing:
\begin{align}
    \frac{c_d}{\sqrt{\Delta_V}} &= c
    \Longrightarrow \Delta_V = (1 - \delta)^2,
\end{align}
which produces the resultant covariance matrix
\begin{align}
    \bs V_d{'} &= \begin{pmatrix} 
        a \id & c \sz \\ 
        c \sz & \LL[ \frac{b_d + 1}{(1 + \delta^2)} - 1 \RR] \id
    \end{pmatrix} \\
    &\equiv \begin{pmatrix} 
        a \id & c \sz \\ 
        c \sz & \LL[ b + T\e_\text{eff} \RR] \id
    \end{pmatrix}. \label{eq:cBenchmarkEquivCV}
\end{align}
However, such a choice is not the optimal strategy. By symmetry, the reduction in correlations associated with the postprocessing operation must also reduce Eve's correlations; similarly, restoring correlations with Alice must also restore Eve's correlations with Bob's measurements. Furthermore, from Eq. \eqref{eq:cBenchmarkEquivCV} we can see that the effective channel associated with setting $c_d' = c$ is the Gaussian excess-noise channel $\mathcal{E}(T, \e + \e_\text{eff})$, where $\e_\text{eff} \simeq 2d^2e_C(1 + 2\delta)$ to first order in $\delta$ (Appendix \ref{app:covar_renorm}). Thus, if Bob sets $\Delta_V = (1 - \delta)^2$ he not only improves Eve's attack but also degrades his effective channel.

Alternatively, Bob may choose instead to apply the amplification operation in such a way that he restores the original variance of his own measurements, i.e.
\begin{align}
    \frac{b_d}{\Delta_V} + \frac{1}{\Delta_V} - 1 &= b \\
    \Longrightarrow \Delta_V = \frac{b_d + 1}{b + 1}.
\end{align}
The covariance matrix then becomes
\begin{align}
    \bs V_d{'} &= \begin{pmatrix} 
        a \id & c\sqrt{\frac{b + 1}{b_d + 1}} \sz \\ 
        c\sqrt{\frac{b + 1}{b_d + 1}} \sz & b \id
    \end{pmatrix} \\
    &\equiv \begin{pmatrix} 
        V \id & \sqrt{T_\text{tot}(V^2 - 1)} \sz \\ 
        \sqrt{T_\text{tot}(V^2 - 1)} \sz & \LL[ T_\text{tot}(V + \e_\text{tot} - 1) + 1 \RR] \id
    \end{pmatrix}, \label{eq:bBenchmarkEquivCV}
\end{align}
i.e. Bob's effective channel is the Gaussian excess-noise channel $\mathcal{E}_\text{eff} = \mathcal{E}(T_\text{tot}, \e_\text{tot})$ (Appendix \ref{app:covar_renorm}). This choice of renormalisation leads to a tighter bound on Eve's information, as it maintains the original fidelity of Bob's quantum signal while also implying a reduction in Eve's ability to estimate Bob's measurement distribution. As long as the total postprocessing and rescaling procedure does not produce a net increase in Alice and Bob's correlations, i.e. the effective channel $\mathcal{E}_\text{eff}$ is physically legitimate (Appendix \ref{app:covar_renorm}), it is sufficient to calculate the Holevo quantity and thus an appropriate bound on the secret-key fraction based on the effective channel. 

Finally, we also observe that estimating the keyrate from the sub-ensemble of a single classical symbol implies the mean of Alice and Bob's distribution remains non-zero after the postprocessing and rescaling operations (Eq. [\ref{eq:dispPrimeMean}]). In practical implementations of the protocol, however, the data may be freely treated as centered \cite{leverrier_composable_2015} without affecting the calculation of the relevant informational quantities from the covariance matrix.
 
\subsection{Asymptotic regime} \label{sec:results_asymptotic}
We now proceed to compute the secret-key generation rate $K^\infty$ of the protocol in the asymptotic ($N \rightarrow \infty$) limit. For each set of free parameters $\{ V, T, \e, d \}$ characterising the protocol, the joint state held by Alice and Bob after heterodyne measurement and postprocessing is represented to Gaussian order by Eqs. \eqref{eq:dispPrimeMean} and \eqref{eq:dispPrimecovar}, from which a lower bound on the distillable secret key fraction can be inferred as per Appendix \ref{app:asym_keyrate}:
\begin{align}
    K^\infty &= \beta I_{AB} - \chi_{EB},
\end{align}
where $\chi_{EB}$ is the Holevo quantity. In accordance with previous works \cite{qi_simultaneous_2016, qi_noise_2018, pan_simultaneous_2020} we calculate the secret-key generation rate under the assumption of a fixed classical quality-of-service: that is, we assume that $d$ is chosen by Alice in such a way as to maintain a classical bit-error rate less than or equal to some threshold $\mathcal{W}$, i.e.
\begin{align}
    \text{BER} &\leq \mathcal{W} \\
    \Longrightarrow \mathcal{W} &\geq \frac{1}{2}\text{erfc} \LL( \frac{\sqrt{T}d}{2\sqrt{T(V + \e - 1) + 2}} \RR),
\end{align}
or equivalently
\begin{align}
    d &\geq 2\hspace{0.1em}\text{erfc}^{-1}(2\mathcal{W}) \sqrt{V + \e - 1 + \frac{2}{T}}.
\end{align}
If the parameters of the protocol are chosen in this way, then for a given distance $L$, channel noise $\e$ and classical quality-of-service $\mathcal{W}$, the only free parameter remaining is Alice's modulation variance $V$. We therefore obtain the maximum lower-bound on the keyrate for fixed $\{L, \e, \mathcal{W} \}$ by optimising over $V$:
\begin{align}
    K &= \max_{V \in [1, \infty]} \ K(V, L, \e, \mathcal{W}).
\end{align}

\begin{figure}[!htb]
    \centering
    \includegraphics[width = \columnwidth]{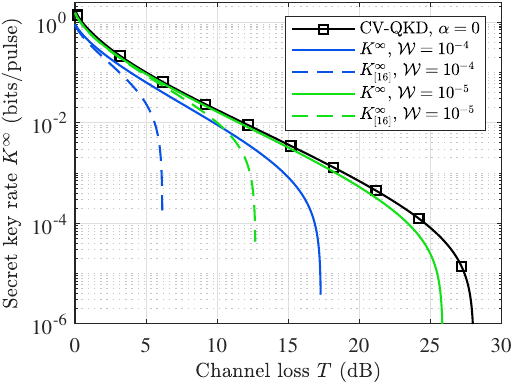}
    \caption{Asymptotic keyrates of the improved SQCC protocol introduced in this work ($K^\infty$, solid lines) compared against the SQCC protocol described by \cite{qi_noise_2018} ($K^\infty_{[16]}$, dashed lines). We assume the untrusted excess noise of the channel is $\e = 0.05$ and the reconciliation efficiency of the protocol is $\beta = 0.95$. At each $T$, the classical displacement $d$ is chosen by Alice such that Bob's classical bit-error rate is fixed at $\mathcal{W}$ and the keyrate is optimised over Alice's modulation variance $V$.}
    \label{fig:pubfig_asymptotic}
\end{figure}

A numerical evaluation of the optimised keyrate is presented in Figure \ref{fig:pubfig_asymptotic}. The performance of the improved model provides similar overall results to previous works, in agreement with common-sense expectations about the performance of the protocol: for no classical signal ($d = 0$, or equivalently $\mathcal{W} = 0.5$), the coupling between the quantum and classical channel vanishes and the optimized secret-key rate becomes equal to the secret-key rate obtained for a heterodyne CV-QKD protocol under the same channel and modulation. As the required maximum bit-error rate threshold decreases, the required classical displacement must increase, increasing the coupling of the two channels towards some maximum value until the classical signal becomes large enough to allow Bob to accurately discern between classical symbols to a level sufficient to not affect his reconstruction of the quantum signal. The coupling thus vanishes as $d$ increases, recovering the original $d = 0$ curve at $d \rightarrow \infty$ or $\mathcal{W} \rightarrow 0$. 

Crucially, however, the new scheme actually presents a significant improvement on the SQCC models used previously in the literature, with our model predicting a positive secure key rate at distances up to double or triple the maximum extent of previous results. Our improved SQCC scheme also approaches the keyrate of the equivalent CV-QKD protocol much more closely for the same required classical quality-of-service than the previous model, which implies our new scheme saturates the upper bound on SQCC protocol performance for substantially lower-energy input states.

As an aside, we note that the classical displacement $d$ required to achieve the desired classical quality-of-service greatly exceeds the relative power of the quantum signal, i.e. $\alpha^2 \geq V + 1$. This large displacement has ramifications for experimental implementation, for example requiring a high dynamic range on the heterodyne detector at Bob's side \cite{qi_simultaneous_2016}, but more importantly it increases the relative significance of phase-noise instability between the signal mode and other modes such as the local oscillator. It is possible to model the effect of errors arising from this phase-noise instability as an additional excess-noise source proportional to the beam power, i.e. $\e_\sigma = \alpha^2 \sigma$ for some constant factor $\sigma$ \cite{qi_noise_2018}; though in the interest of elucidating only the effect of the quantum-classical coupling we have elected to not include it in the presented calculations. However, preliminary results assuming the inclusion of phase noise $\sigma = 10^{-4}$ largely do not change the overall relative behaviour of the protocols shown in Figs. \ref{fig:pubfig_asymptotic} and \ref{fig:pubfig_finitesize}, with the new scheme continuing to provide a clear advantage. This is expected: since the improved scheme reduces the relative classical power required to achieve an equivalent quantum quality-of-service, the effect of the power-dependent phase noise contribution is also reduced in comparison to the original model.

\subsection{Finite-key regime} \label{sec:results_finitesize}
Assessing the security of the protocol for finite block sizes requires a slightly more sophisticated analysis. We utilise the results presented in Ref. \cite{jain_practical_2022} to estimate a lower bound on the achievable secret-key fraction of the protocol under the requirement of composable security, which states that a Gaussian-modulated coherent-state QKD protocol of block size $N$ and frame-error reconciliation rate $p_f$ is $\epsilon$-secure for
\begin{align}
    \epsilon &= \epsilon_\text{qrng}+\epsilon_h+\epsilon_s+\epsilon_\text{IR}+\epsilon_\text{ent}+\epsilon_\text{PE}+\epsilon_\text{cal},
\end{align}
conditional on the key length $\ell$ being chosen such that
\begin{align}
    \ell \leq &\ p_fNK_\text{PE}^\infty - \sqrt{p_fN}\Delta_\text{AEP} \notag \\
    &- \sqrt{p_fN}\log_2(p_fN)\Delta_\text{ent} + \Delta_S + \Delta_H.
\end{align}
In the above expression, each security parameter $\epsilon_k$ corresponds to a confidence probability in performing a particular step of the protocol; for example, $\epsilon_{PE}$ is the likelihood that Eve's information is greater than the lower bound given by the parameter estimation process. (A further description of each $\epsilon_k$ is given in Appendix \ref{app:fnsz_keyrate}, Table \ref{tab:security_params}.) The probability of the protocol failing is therefore given by $\epsilon$. The corrective terms $\Delta_k$ account for information leakage as a result of bounding the conditional min-entropy ($\Delta_\text{AEP}$), conditional von Neumann entropy ($\Delta_\text{ent}$) and smooth min-entropy after frame reconciliation ($\Delta_S$), as well as information loss occurring from the privacy amplification procedure as per the leftover hash lemma \cite{tomamichel_framework_2012} ($\Delta_H$):
\begin{align}
    \Delta_\text{AEP} &= 4(d_{rx} + 1) \sqrt{\log_2\LL( \frac{2}{(\frac{p_f}{3} \epsilon_s^2)^2} \RR)} \\
    \Delta_\text{ent} &= \sqrt{2 \log_2 \LL( \frac{2}{\epsilon_\text{ent}} \RR)} \\
    \Delta_S &= \log_2 \LL( p_f - \frac{p_f \epsilon_s^2}{3} \RR) \\
    \Delta_H &= 2\log_2\big( \sqrt{2}\epsilon_h \big).
\end{align}
The quantity $K_\text{PE}^\infty$ represents the asymptotic keyrate calculated according to the worst-case parameter estimates for block size $N$ with confidence level $\epsilon_\text{PE}$:
\begin{align}
    K_\text{PE}^\infty &= \beta \overline{I}(\hat \rho_{G}^{AB}) - f(\Sigma_a^{max}, \Sigma_b^{max}, \Sigma_c^{min}).
\end{align}
Further technical details on the parameter estimation process are presented in Appendix \ref{app:fnsz_keyrate}. The maximum secret-key bits shared per pulse $K^\mathcal{F}~=~\ell/N$ of such a protocol is therefore bounded below by
\begin{align}
    K^\mathcal{F} \geq &\ p_f \LL[ \beta\overline{I}\LL( \hat \rho_G^{AB} \RR) - f(\Sigma_a^{max}, \Sigma_b^{max}, \Sigma_c^{min}) \RR] \notag\\
    &- \sqrt{\frac{p_f}{N}}\Delta_\text{AEP} - \sqrt{\frac{p_f\log_2(p_fN)}{N}}\Delta_\text{ent} \notag\\
    &+ \frac{\Delta_S}{N} + \frac{\Delta_H}{N}.
\end{align}
For ease, we choose our protocol and security parameters largely in accordance with Ref. \cite{jain_practical_2022}. These are summarised in Appendix \ref{app:fnsz_keyrate}, Table \ref{tab:security_params}.

\begin{figure}[!tb]
    \centering
    \includegraphics[width = \columnwidth]{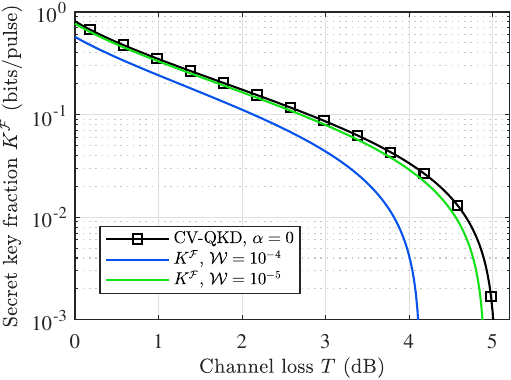}
    \caption{Finite-size secret-key fractions for the improved SQCC protocol ($K^\mathcal{F}$, solid lines). The secret-key fraction is calculated from initial block size $N = 10^8$ under the same protocol parameters and optimization as in Fig. \ref{fig:pubfig_asymptotic}. Further finite-size security parameters are detailed in Appendix \ref{app:fnsz_keyrate}, Table \ref{tab:security_params}.}
    \label{fig:pubfig_finitesize}
\end{figure}

Special attention should be paid here to the order in which the protocol must be conducted, such that no step of the protocol requires information Alice and Bob do not already have. For example, Alice and Bob must perform the parameter estimation step after Bob postprocesses and rescales his data points, since it is not known how to estimate the channel characteristics from the unprocessed four-mode mixed Gaussian distribution that Bob measures. We propose the following scheme:
\begin{enumerate}
    \item Bob performs heterodyne measurement on their shared states.
    \item Bob estimates from his raw data the peak location of each of his four Gaussian modes. This informs his choice of magnitude when performing the re-displacement operation.
    \item Bob postprocesses his data by re-displacing his points according to a threshold discrimination.
    \item Bob rescales his data points to renormalise the distribution. In order to do this, Bob must determine the value $\Delta_V = (b_d + 1)/(b + 1)$, which requires him to properly estimate the original variance of his data, $b$, from the variance of his postprocessed measurements $b_d$. Bob must therefore estimate the signal-to-noise ratio, which subsequently allows him to estimate $e_C$ and $\delta$ and so infer the value of $b$. Alice and Bob do this by publicly comparing a fraction of their classical results and evaluating an upper bound on the bit-error rate, which allows them to determine the signal-to-noise ratio.
    \item Alice and Bob perform error correction on the postprocessed and renormalised data.
    \item Alice and Bob perform parameter estimation on their corrected postprocessed and renormalised data, assuming knowledge of only the first and second statistical modes.
    \item Alice and Bob distil a secret key via privacy amplification.
\end{enumerate}
We note in Steps 2 and 4 that such an estimation is equivalent to performing additional parameter estimation steps, on Alice's mean value parameter $\alpha = \sqrt{T}d$ and the signal-to-noise quantity SNR respectively. When computing a lower bound on Eve's information, Bob must therefore take into account these additional estimations by evaluating the Holevo quantity using the worst-case estimators of the two parameters. However, we predict the effect of these additional parameter estimation steps to be minimal in any practical regime.

The numerical results associated with the finite-key case are presented in Figure \ref{fig:pubfig_finitesize}, where we have also performed the same procedure of optimisation for a fixed classical threshold as described in Section \ref{sec:results_asymptotic}. Our scheme successfully generates a positive secure key fraction in the finite-key regime, even at modest block sizes ($N = 10^{8}$). As might be expected, the secure key fraction generated in the finite-size regime largely reproduces the trends of the asymptotic case, demonstrating a clear advantage over the previous SQCC scheme, though to a lesser degree.

\section{Discussion} \label{sec:discussion}
In this work, we present an updated and improved model of simultaneous quantum-classical communication in Gaussian-modulated coherent-state CV-QKD which addresses gaps in previous analyses. We present a more sophisticated analysis of the coupling between the classical and quantum channels in the SQCC protocol, and discuss differences with previous works \cite{qi_noise_2018}. Evidence is provided for the proposed model via Monte Carlo simulation analysis and we observe excellent agreement between analytic prediction and simulated results. We develop a comprehensive framework to address potential security issues inherently associated with SQCC communication and show that our proposed model is not only more accurate, but also presents improved quantum performance for a fixed classical quality-of-service. We compute numerical results showing the secret-key generation rate of the SQCC protocol for different classical performance thresholds in the asymptotic regime and extend our analysis to the finite-key regime using contemporary finite-key bounds. Lastly, we have demonstrated a secure implementation of the SQCC protocol under realistic conditions in the finite-key case, something that has not previously been described. Our results thus represent an invaluable step towards the implementation of hybrid quantum-classical communications protocols. Further improvements could also be obtained by forward error-correcting the classical signal, increasing the effectiveness of both the quantum and classical aspects of the protocol by increasing the effective signal-to-noise ratio, at the cost of classical data throughput.

While the results presented in this work are essentially fundamental and mostly implementation- and platform-agnostic, certain assumptions, e.g. fixed channel loss, are not necessarily true for all applications. For example, we anticipate that SQCC protocols may have an advantage on satellite-based platforms, where size and power usage are heavily economised. However, satellite uplink (Earth-to-satellite) or downlink (satellite-to-Earth) configurations may suffer from fading channels, which we do not consider in the main body of this work; in Appendix E we briefly discuss additional issues that arise for SQCC under such conditions.

\begin{acknowledgments}
The Australian Government supported this research through the Australian Research Council’s Linkage Projects funding scheme (Project No. LP200100601). The views expressed herein are those of the authors and are not necessarily those of the Australian Government or the Australian Research Council. Approved for Public Release: NG25-0640.
\end{acknowledgments}

\appendix
\section{Husimi Q-function formulation} \label{app:qfunction}
Since the object of this work is to model the effect of one or more postprocessing operations on a continuous-variable protocol using heterodyne measurements, it is advantageous to describe our quantum states through the Husimi Q-function representation. The Q-function of an $n$-mode quantum state $\hat \rho$ is defined in the following way \cite{zachos_quantum_2005}:
\begin{align}
    Q(\bs{\alpha}) &= \frac{1}{\pi^n}\bra{\bs{\alpha}}\hat \rho\ket{\bs{\alpha}},
\end{align}
where $\bs{\alpha} \in \mathbb{C}^n$ and $\ket{\bs{\alpha}} \equiv \bigotimes_{i=0}^n \ket{\alpha_i}$. By construction, the Q-function is the normalised and positive probability density function \cite{cartwright_non-negative_1976} associated with projection onto a specific coherent state $\ket{\bs{\alpha}}$, which also implies it is the probability density function associated with Alice and Bob's measurement results for their protocol when using heterodyne measurement \cite{weedbrook_gaussian_2012}.

Since our protocol deals commonly with Gaussian states, it is useful to derive a general formula for the Q-function of a Gaussian state given an arbitrary mean vector and covariance matrix. Using shot-noise units ($\hbar = 2$) \cite{laudenbach_continuousvariable_2018}, the Wigner function of a general $n$-mode Gaussian state is given by \cite{weedbrook_gaussian_2012}
\begin{align}
    W^{\bs \mu}_{\bs V}(\bs \xi) &= \frac{e^{-\frac{1}{2}(\bs \xi - \bs \mu)^T\bs{V}^{-1}(\bs{\xi} - \bs{\mu})}}{(2\pi)^n \sqrt{\det(\bs V)}}
\end{align}
for ${\bs{\mu}} = (\langle\hat q_A\rangle, \langle\hat p_A\rangle, ...)^T$ the mean of the quantum state, $\bs \xi \in \mathbb{R}^{2n}$ and $\bs V \in \text{Sp}(2n, \mathbb{R})$ the symplectic covariance matrix. We obtain the Husimi Q-function from the Wigner function via Weierstrass transform:
\begin{align}
    Q(\bs \alpha) &= \bigg( \frac{2}{\pi} \bigg)^n \int \de\bs \xi \ W(\bs \xi) e^{-\frac{1}{2}(2\bs \alpha - \bs \xi)^T(2\bs \alpha - \bs \xi)}.
\end{align}
It is not difficult to show then that the generalised Q-function of a Gaussian state with mean vector $\bs{\mu}$ and covariance matrix $\bs{V}$ is given by
\begin{align}
    Q^{\bs \mu}_{\bs V}(\bs \alpha) = \frac{\sqrt{\det(\bs M)}}{\pi^N} e^{-(\bs \alpha - \frac{\bs \mu}{2})^T \bs M (\bs \alpha - \frac{\bs \mu}{2})},
\end{align}
for $\bs{M} = 2(\bs{V} + \id)^{-1}$, when $V$ has the general form given by Eq. \eqref{eq:init_covar}. (This is consistent with the zero-mean expression given in \cite{hosseinidehaj_finite-size_2020}.) Finally, the statistical moments of the quadrature operators in the quantum space are functions of the statistical moments of the Q-function in the phase space:
\begin{align}
    \langle \hat q_i \rangle &= \int \de^n\bs{\alpha} \ 2\Re[\alpha_i] \cdot Q(\bs{\alpha}) \label{eq:qfunc_m}, \\
    \langle \hat q_i^2 \rangle &= \int \de^n\bs{\alpha} \ (4\Re[\alpha_i]^2 - 1) Q(\bs{\alpha}) \label{eq:qfunc_cv1}, \\
    V_{\hat q_i, \hat q_i} &= \langle \hat q_i^2 \rangle - \langle \hat q_i \rangle^2, \ \text{etc}. \label{eq:qfunc_cv2}
\end{align}
A useful property of the Husimi Q-function is that for a Gaussian quantum state of mean $\bs \mu$ and covariance $\bs V$, the Q-function is itself Gaussian in $\bs \alpha$ with mean $\bs \mu/2$ and covariance $(\bs V + \id)/4$.

\section{Derivation of the covariance matrix} \label{app:covar}
As discussed in Section \ref{sec:protocol_description_specific}, we bound the security of the protocol by exploiting Gaussian optimality \cite{navascues_optimality_2006, garcia-patron_unconditional_2006} and extremality \cite{wolf_extremality_2006} results to estimate upper and lower bounds on the  Holevo quantity and mutual information respectively. If Bob and Alice share some jointly correlated measurements $\{\tilde \alpha_i, \tilde \beta_i\}$, which are rendered non-Gaussian by some postprocessing operation, then the true secret key fraction distillable from $\{\tilde \alpha_i, \tilde \beta_i\}$ is bounded below by the key fraction calculated for an equivalent Gaussian protocol in which Alice and Bob share a Gaussian state $\hat \rho_G$ with the same first- and second-order quadrature moments as $\{\tilde \alpha_i, \tilde \beta_i\}$. Hence it suffices for our security analysis to calculate only the equivalent quadrature mean vector and covariance matrix of the postprocessed and rescaled joint measurements.

Initially, Alice and Bob start with an ordinary Gaussian-modulated coherent-state protocol using heterodyne detection \cite{weedbrook_quantum_2004} over a lossy bosonic channel $\mathcal{E}$ with transmissivity $T$ and thermal-noise contribution $\e$, over which they share $n$ individual shots. For each shot, Alice performs the displacement operation described in Section \ref{sec:protocol_description_specific} to transmit an additional classical symbol of amplitude $\sqrt{T}d$ to Bob. In the entanglement-based picture, Alice and Bob now jointly possess a bivariate Gaussian distribution of measurement results $\{\alpha_1, ... \alpha_n, \beta_1, ... \beta_n\}$ with equivalent quadrature operator statistics \cite{laudenbach_continuousvariable_2018}
\begin{align}
    \bs{\mu} &= \LL( 0, 0, \frac{\sqrt{T} d}{\sqrt{2}}, \frac{\sqrt{T} d}{\sqrt{2}}\RR)^T \label{eq:mAB}, \\
    \bs{V} &= \begin{pmatrix}
        V \id & \sqrt{T(V^2 - 1)} \sz \\
        \sqrt{T(V^2 - 1)} \sz & \LL[T(V + \e - 1) + 1\RR] \id
    \end{pmatrix} \\
    &= \begin{pmatrix}
        a \id & c \sz \\
        c \sz & b \id
    \end{pmatrix} \label{eq:cvAB}.
\end{align}
Recall that the security of the total SQCC protocol is inferred by estimating the security for the sub-ensemble of only a single classical symbol; for ease, we have chosen the ensemble where Alice sends the symbol
\begin{align}
    \tilde d = d_1 = \frac{d}{\sqrt{2}} + i\frac{d}{\sqrt{2}},
\end{align}
though this choice is arbitrary and will not affect the final security estimation.

\subsection{Measurement statistics after postprocessing} \label{app:covar_postprocessing}
In order to decompose the combined signal into separate classical and quantum data streams, Bob follows the threshold discrimination process described in Section \ref{sec:protocol_description_general} to firstly identify the each sent classical symbol and secondly emulate Alice's zero-mean Gaussian distribution by inverting the displacement. We can codify this process using the formalism introduced in Appendix \ref{app:qfunction}.

Because Alice and Bob use heterodyne detection, the probability distribution of their joint data is given by the Husimi Q-function parameterised by the mean vector \eqref{eq:mAB} and covariance matrix \eqref{eq:cvAB}. (We use the shorthand $Q^{z}$ to refer to a Q-function with covariance matrix \eqref{eq:cvAB} and mean vector $\bs{\mu} = (0, 0, \sqrt{T}\Re\{\tilde z\}, \sqrt{T}\Im\{\tilde z\})^T$.) The Q-function describing the initial measurement distribution is therefore
\begin{align}
    Q_{in} &= Q^{d_1}(\tilde \alpha, \tilde \beta).
\end{align}
Bob separates his data according to position in the four quadrants of the complex plane and assigns the appropriate classical symbol to each one:
\begin{align}
    Q_{in} = \begin{cases} 
    Q^{d_1}(\tilde \alpha, \tilde \beta)   
    & \beta_x\geq 0, \beta_y \geq 0 \ (d_1, 00) \\
    Q^{d_1}(\tilde \alpha, \tilde \beta)     
    & \beta_x < 0, \beta_y > 0 \ (d_2, 01)  \\
    Q^{d_1}(\tilde \alpha, \tilde \beta)       
    & \beta_x \leq 0, \beta_y \leq 0 \ (d_3, 11)  \\
    Q^{d_1}(\tilde \alpha, \tilde \beta)      
    & \beta_x > 0, \beta_y < 0  \ (d_4, 10).  \\
   \end{cases}
\end{align}
He then virtually performs the re-displacement operation $\hat D(-\sqrt{T}\tilde d)$ to each shot according to the measured classical symbol to return the distribution to a zero-mean one, approximately reproducing the original quantum signal. This is equivalent to the transform $\tilde \beta \Longrightarrow \tilde \beta' = \tilde \beta - \sqrt{T}\tilde d$, and so the Q-function $Q_d$ describing the distribution of Alice and Bob's re-displaced quantum measurement results becomes
\begin{align}
    Q_d &= \begin{cases} 
        Q^{d_1}(\tilde \alpha, \tilde \beta' + \sqrt{T}d_1)
        & \beta_x'+\frac{\sqrt{T}d}{2\sqrt{2}}\geq 0, \beta_y'+\frac{\sqrt{T}d}{2\sqrt{2}} \geq 0 \\
        Q^{d_1}(\tilde \alpha, \tilde \beta' + \sqrt{T}d_2)
        & \beta_x'-\frac{\sqrt{T}d}{2\sqrt{2}}< 0, \beta_y'+\frac{\sqrt{T}d}{2\sqrt{2}} > 0 \\
        Q^{d_1}(\tilde \alpha, \tilde \beta' + \sqrt{T}d_3)
        & \beta_x'-\frac{\sqrt{T}d}{2\sqrt{2}}\leq 0, \beta_y'-\frac{\sqrt{T}d}{2\sqrt{2}} \leq 0 \\
        Q^{d_1}(\tilde \alpha, \tilde \beta' + \sqrt{T}d_4)
        & \beta_x'+\frac{\sqrt{T}d}{2\sqrt{2}}> 0, \beta_y'-\frac{\sqrt{T}d}{2\sqrt{2}} < 0 \\
    \end{cases}
\end{align}
\begin{align}
    &= \begin{cases} 
        Q^{0}(\tilde \alpha, \tilde \beta')
        & \beta_x' \geq -\frac{\sqrt{T}d}{2\sqrt{2}}, \ \beta_y' \geq -\frac{\sqrt{T}d}{2\sqrt{2}} \\
        Q^{d_1-d_2}(\tilde \alpha, \tilde \beta')
        & \beta_x' < \frac{\sqrt{T}d}{2\sqrt{2}}, \ \ \ \beta_y' > -\frac{\sqrt{T}d}{2\sqrt{2}} \\
        Q^{d_1-d_3}(\tilde \alpha, \tilde \beta')
        & \beta_x' \leq \frac{\sqrt{T}d}{2\sqrt{2}}, \ \ \ \beta_y' \leq \frac{\sqrt{T}d}{2\sqrt{2}} \\
        Q^{d_1-d_4}(\tilde \alpha, \tilde \beta')
        & \beta_x' > -\frac{\sqrt{T}d}{2\sqrt{2}}, \ \beta_y' < \frac{\sqrt{T}d}{2\sqrt{2}} \\
    \end{cases}\\
    &\equiv \begin{cases} 
        Q^{0}
        & \beta_x \geq -\frac{\sqrt{T}d}{2\sqrt{2}}, \ \beta_y \geq -\frac{\sqrt{T}d}{2\sqrt{2}} \\
        Q^{2 d/\sqrt{2}}
        & \beta_x < \frac{\sqrt{T}d}{2\sqrt{2}}, \ \ \ \beta_y > -\frac{\sqrt{T}d}{2\sqrt{2}} \\
        Q^{2 d/\sqrt{2}+2i d/\sqrt{2}}
        & \beta_x \leq \frac{\sqrt{T}d}{2\sqrt{2}}, \ \ \ \beta_y \leq \frac{\sqrt{T}d}{2\sqrt{2}} \\
        Q^{2i d/\sqrt{2}}
        & \beta_x > -\frac{\sqrt{T}d}{2\sqrt{2}}, \ \beta_y < \frac{\sqrt{T}d}{2\sqrt{2}}. \\
    \end{cases} \label{eq:qDisp}
\end{align}
It is clear to see from Eq. \eqref{eq:qDisp} how the necessarily-imperfect classical signal readout effectively couples the classical channel with the quantum channel. For $d = 0$, i.e. no classical signal, Bob's distribution reduces down to $Q^0$ (Alice's original zero-mean distribution). However, for $d \neq 0$ Bob's distribution only approximates $Q^0$, as the distribution gains additional non-Gaussian corrections from measurements which were incorrectly identified and displaced in the wrong direction. Moreover, as $\sqrt{T}d$ becomes much larger than the variance of the quantum signal, Bob's classical bit errors become fewer and the subsequent corrections reduce in size until Bob's distribution again approaches $Q^0$, as expected.

After postprocessing, it is straightforward to calculate the Gaussian-order statistics of Bob's new distribution $Q_d$ via Eqs. \eqref{eq:qfunc_m}-\eqref{eq:qfunc_cv2}:
\begin{align}
    \bs{\mu}_d &= \LL( 0, 0, \frac{2\sqrt{T} d e_C}{\sqrt{2}}, \frac{2\sqrt{T} d e_C}{\sqrt{2}} \RR)^T, \\
    \bs{V}_d &= \begin{pmatrix}
        a_d\id & c_d\sz \\
        c_d\sz & b_d\id
    \end{pmatrix}
\end{align}
where $e_C$ is the classical bit-error rate
\begin{align}
    e_C &= \frac{1}{2}\text{erfc}\left( \frac{ \sqrt{T} d}{2\sqrt{b+1}} \right)
\end{align}
and
\begin{align}
    a_d &= a, \\
    b_d &= b - \sqrt{\frac{4(b+1)}{\pi}}  \sqrt{T} d e^{-\frac{ T d^2}{4(b+1)}} \notag \\
    &\quad + 2T d^2e_C - 2 T d^2e_C^2, \\
    c_d &= c \left( 1 - \frac{ \sqrt{T} d e^{-\frac{ T d^2}{4(b+1)}}}{\sqrt{\pi(b+1)}} \right).
\end{align}
We can simplify the analytic expression considerably by defining the signal-to-noise parameter
\begin{align}
    \text{SNR} &= \frac{T d^2}{b+1},
\end{align}
which implies 
\begin{align}
    e_C &= \frac{1}{2}\text{erfc} \left( \frac{\sqrt{T} d}{2\sqrt{b+1}} \right)\\
    &\equiv \frac{1}{2}\text{erfc} \left( \frac{\sqrt{\text{SNR}}}{2} \right).
\end{align}
Defining also
\begin{align}
    \delta &= \frac{\sqrt{T} d e^{-\frac{T d^2}{4(b+1)}}}{\sqrt{\pi(b+1)}}\\
    &\equiv \sqrt{\frac{SNR}{\pi}} e^{-\frac{SNR}{4}},
\end{align}
we see that
\begin{align}
    a_d &= a, \label{eq:aDisp}\\
    b_d &= b + 2 T d^2 e_C - 2(b+1)\delta - 2 T d^2e_C^2, \label{eq:bDisp}\\
    c_d &= c \left( 1 - \delta \right). \label{eq:cDisp}
\end{align}
In addition, the postprocessed state $Q_d$ remains normalised, i.e. $||Q_d|| = 1$. This is to be expected, as we are not removing any data points or altering the probability `volume' using any methods other than slicing and translation, which leaves the total probability the same as the initial (normalised) state.

\subsection{Renormalisation} \label{app:covar_renorm}
As with other protocols involving postprocessing, it is necessary for Bob to `physicalise' his data after performing the re-displacement operation. Such a step is required to guarantee the existence of the effective Gaussian state $\hat \rho_G$ on which Gaussian optimality and extremality is predicated, as well as to guarantee that Eve can legitimately hold a purification of Alice and Bob's joint state \cite{laudenbach_continuousvariable_2018, pirandola_advances_2020}, which is necessary for Alice and Bob to infer the Holevo quantity from their data. While the norm of the state is already unity, Eq. \eqref{eq:bDisp} shows that for certain conditions it is possible for Bob to simultaneously obtain effective variances below the shot noise (i.e. $b_d < 1$), something that is obviously incompatible with any physical state or channel. 

\begin{figure}[!htb]
    \centering
    \begin{tikzpicture}[
    RDET/.style={
    semicircle, draw=black!80, fill=gray!40, thick, minimum size=3mm, rotate = 270
    },
    LABEL/.style={
    rectangle
    },
    OPERATOR/.style={
    rectangle, rounded corners, draw=black!80, fill=blue!20, thick, minimum size=7.5mm
    },
    POSTDISPOP/.style={
    rectangle, rounded corners, draw=black!80, dashed, fill=gray!10, thick, minimum size=7.5mm
    }
    ]
    \node[LABEL](origin2) at (0,1.75) {};
    \node[RDET](rDet2) [left = 0.375cm of origin2.center, anchor = chord center]{};
    \node[POSTDISPOP](ppOperator2)[right = 0.5cm of rDet2.north, anchor = west]{PP};
    \node[POSTDISPOP](rsOperator2)[right = 0.7cm of ppOperator2.east, anchor = west]{RS};
    \node[LABEL](incomingQ2) at (-2, 1.75) {};
    \node[LABEL](outgoingQ2) at (4, 1.75) {};
    
    \node[LABEL](incomingQLabel2) at (-1.125, 1.35) {$\hat q$};
    \node[LABEL](ppQLabel2) at (1.6, 1.35) {$\hat q_d$};
    \node[LABEL](rsQLabel2) at (3.25, 1.35) {$\hat q_d'$};
    
    \draw[->] (incomingQ2) -- (rDet2.chord center);
    \draw[decorate,decoration={coil}] (rDet2.north) -- (ppOperator2.west);
    \draw[decorate,decoration={coil,aspect=0}] (ppOperator2.east) -- (rsOperator2.west);
    \draw[decorate,decoration={coil,aspect=0}] (rsOperator2.east) -- (outgoingQ2);

    \node[LABEL](origin) at (0,0) {};
    \node[OPERATOR](pOperator) at (0.75,0) {$\mathcal{\hat P}$};
    \node[RDET](rDet)[right = 0.5cm of pOperator.east, anchor = chord center]{};
    \node[LABEL](incomingQ) at (-2, 0) {};
    \node[LABEL](outgoingQ) at (4, 0) {};
    
    \node[LABEL](incomingQLabel) at (-1.125, 0.4) {$\hat q$};
    \node[LABEL](outgoingQLabel) at (3.25, 0.4) {$\hat q'$};
    
    \draw[-] (incomingQ) -- (pOperator.west);
    \draw[->] (pOperator.east) -- (rDet.chord center);
    \draw[decorate,decoration={coil,aspect=0}] (rDet.north) -- (outgoingQ);
    
\end{tikzpicture}
    \caption{The physical protocol (top) and equivalent virtual protocol (bottom) of the SQCC scheme. For an inferred keyrate to be a valid lower bound on the distillable secret information, the effective virtual channel $\mathcal{\hat P}$ that emulates the postprocessing and rescaling must be identifiable with a physically legitimate operation.}
    \label{fig:physicality_diagram}
\end{figure}
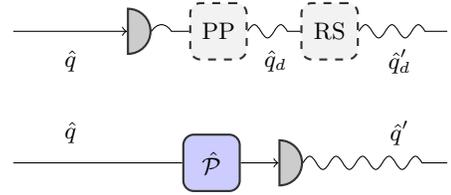

We choose to approach the issue of physicalisation by implementing a rescaling of the postprocessed data points 
\begin{align}
    \tilde \beta_d \longrightarrow \tilde \beta_d' = \frac{\tilde \beta_d}{\sqrt{\Delta_V}},
\end{align}
which is equivalent to an electronic amplification of gain $G = 1/{\sqrt{\Delta_V}}$ \cite{bachor_guide_2019}. The physicality requirement then becomes a choice on $\Delta_V$ such that the combined postprocessing and rescaling operations virtually emulate some equivalent physical operation $\mathcal{\hat P}$ (Figure~\ref{fig:physicality_diagram}).

Analytically, Figure \ref{fig:physicality_diagram} implies the total effect of the two virtual operations performed by Bob
\begin{align}
    \begin{pmatrix}
        a\id & c\sz \\
        c\sz & b\id
    \end{pmatrix} \rightarrow \begin{pmatrix}
        a\id & \frac{c_d}{\sqrt{\Delta_V}}\sz \\
        \frac{c_d}{\sqrt{\Delta_V}}\sz & \LL[ \frac{b_d}{\Delta_V} + \frac{1}{\Delta_V} - 1 \RR]\id
    \end{pmatrix} \label{eq:postprocessing_rescaling_transform}
\end{align}
should be equivalent to the action of $\mathcal{\hat P}$ on the joint covariance matrix 
\begin{align}
    \hat q_0 &\rightarrow \mathcal{\hat P}^\dagger \hat q_0 \mathcal{\hat P} \\
    \Longrightarrow \begin{pmatrix}
        a\id & c\sz \\
        c\sz & b\id
    \end{pmatrix} &\rightarrow \begin{pmatrix}
        a'\id & c'\sz \\
        c'\sz & b'\id
    \end{pmatrix}.
\end{align}
We are given some freedom in the choice of our normalisation. Recalling that $c_d = (1 - \delta) c$ for $0 \leq \delta < 1$, one potential option is defining
\begin{align}
    \Delta_V &= (1 - \delta)^2,
\end{align}
such that the net effect of the postprocessing and rescaling does not affect the correlations between Alice and Bob's measurements. Eq. \eqref{eq:postprocessing_rescaling_transform} thus becomes
\begin{align}
    \begin{pmatrix}
        a\id & c\sz \\
        c\sz & b\id
    \end{pmatrix} \rightarrow \begin{pmatrix}
        a\id & c\sz \\
        c\sz & \LL[ \frac{b_d}{(1 - \delta)^2} + \frac{1}{(1 - \delta)^2} - 1 \RR]\id
    \end{pmatrix}.
\end{align}
As long as we maintain the inequality
\begin{align}
    \frac{b_d}{(1 - \delta)^2} + \frac{1}{(1 - \delta)^2} - 1 \geq b,
\end{align}
then by defining 
\begin{align}
    T\e_\text{eff} &= \LL( \frac{b_d}{(1 - \delta)^2} + \frac{1}{(1 - \delta)^2} - 1 \RR) - b
\end{align}
we can simplify Eq. \eqref{eq:postprocessing_rescaling_transform} further:
\begin{align}
    \begin{pmatrix}
        a\id & c\sz \\
        c\sz & b\id
    \end{pmatrix} \rightarrow \begin{pmatrix}
        a\id & c\sz \\
        c\sz & \LL( b + T\e_\text{eff} \RR)\id
    \end{pmatrix} \label{eq:postprocessing_rescaling_transform_simplified}.
\end{align}
It is thus straightforward to identify the effective physical transformation $\mathcal{\hat P}$ equivalent to Eq. \eqref{eq:postprocessing_rescaling_transform_simplified} simply as an additional thermal-noise source with noise parameter $\e_\text{eff}$, or alternatively that Alice and Bob exchange their states over the effective channel $\mathcal{E}(T, \e + \e_\text{eff})$. The physicality of the protocol is therefore guaranteed, and Alice and Bob may safely infer the security of the protocol from their postprocessed and rescaled data with joint Gaussian statistics given by
\begin{align}
    \bs{\mu}_d' &= \LL( 0, 0, \frac{2\sqrt{T} d e_C}{\sqrt{2(1 - \delta)^2}}, \frac{2\sqrt{T} d e_C}{\sqrt{2(1 - \delta)^2}} \RR)^T, \label{eq:mABdispP} \\
    \bs{V}_d' &= \begin{pmatrix}
        a\id & c\sz \\
        c\sz & \LL[ \frac{b_d + 1}{(1 - \delta)^2} - 1 \RR]\id
    \end{pmatrix}. \label{eq:cvABdispP}
\end{align}
The above expression can be simplified further by expanding in terms of $\delta$ and restricting to first order in $\delta$ and $e_C$, since $\delta, e_C \ll 1$ for most practical choices of $d$:
\begin{align}
    &\frac{b_d + 1}{(1 - \delta)^2} - 1 \notag\\
    &= \frac{b + 2 T d^2 e_C - 2(b+1)\delta + 1}{(1 - \delta)^2} - 1 \notag\\
    &= \LL( b + 2 T d^2 e_C - 2b\delta - 2\delta + 1 \RR)\LL[ 1 + 2\delta + O(\delta^2) \RR] - 1 \notag\\
    &= b + 2 T d^2e_C - 2b\delta - 2\delta + 1 \notag\\
    &\quad + 2\delta b + 4\delta T d^2e_C - 4b\delta^2 - 4\delta^2 + 2\delta - 1 + 
    O(\delta^2) \notag\\
    &\simeq b + 2 T d^2 e_C (1 + 2\delta).
\end{align}
Consequently, the covariance matrix \eqref{eq:cvABdispP} under this choice of renormalisation appears strikingly similar to the covariance matrix presented by Qi et al \cite{qi_noise_2018}, where the classical coupling to the quantum channel is proposed to be an additional untrusted thermal-noise source $\e_\text{eff} = 4 d^2 e_C$. This leads to our new scheme providing virtually no real advantage or disadvantage in the keyrate compared to previous literature.

However, the specific choice of renormalisation $\Delta_V~=~(1 - \delta)^2$, such that $c' = c$, is by no means the optimal choice. An alternative solution might be to choose $\Delta_V$ such that Bob's covariance matrix element is invariant under the postprocessing and rescaling, i.e.
\begin{align}
    \frac{b_d}{\Delta_V} + \frac{1}{\Delta_V} - 1 &= b \\
    \Longrightarrow \Delta_V &= \frac{b_d + 1}{b + 1}.
\end{align}
Such a choice is intuitively more reasonable than defining $c' = c$: the net effect of the postprocessing operation is to scale the correlation term $c$ between Alice and Bob by the factor $1-\delta$, and so by extension the effect of the postprocessing must also scale the correlations between Eve's measurements and Bob's by the same factor. Thus, rescaling such that $\sqrt{\Delta_V} = 1 - \delta$ implies that Bob is not only restoring his correlations with Alice's measurements but also with Eve's, increasing Eve's information. Moreover, the rescaling overestimates the effective noise contribution to the channel, further decreasing the keyrate. However, setting $b' = b$ allows Bob to incur no extra effective excess noise contribution and additionally reduce his correlations with Eve by some amount.

The net result of the two virtual operations then becomes
\begin{align}
    \begin{pmatrix}
        a\id & c\sz \\
        c\sz & b\id
    \end{pmatrix} \rightarrow \begin{pmatrix}
        a\id & c_d\sqrt{\frac{b + 1}{b_d + 1}}\sz \\
        c_d\sqrt{\frac{b + 1}{b_d + 1}}\sz & b\id
    \end{pmatrix}. \label{eq:bBenchmarkCV}
\end{align}
Eq. \eqref{eq:bBenchmarkCV} describes a physical operation as long as Alice and Bob observe no net increase in the correlation term, i.e.
\begin{align}
    c' &= \frac{c_d}{\sqrt{\Delta_V}} = c_d\sqrt{\frac{b + 1}{b_d + 1}} \leq c, \label{eq:bBenchmark_physicalityinequality}
\end{align}
since generation of increased entanglement deterministically via local operations is forbidden by quantum mechanics \cite{nielsen_quantum_2010}.

The effective channel or operation $\mathcal{\hat P}$ is slightly more difficult to identify for the transformation represented by Eq. \eqref{eq:bBenchmarkCV}. However, let us consider the covariance matrix of a signal sent consecutively through not only the original lossy bosonic channel $\mathcal{E}(T, \e)$ but also some additional lossy channel $\mathcal{E}_v(T_v, \e_v)$. The covariance matrix of the shared state in such an instance is given by
\begin{align}
    V &= \begin{pmatrix}
        V\id & \sqrt{T T_v(V^2 - 1)}\sz \\
        \sqrt{T T_v(V^2 - 1)}\sz & \LL[ T T_v(V + \e + \frac{\e_v}{T} - 1) + 1 \RR]\id
    \end{pmatrix}. \label{eq:twochannelCV}
\end{align}
If we define
\begin{align}
    T_v &= \frac{c'}{c} = (1-\delta)\sqrt{\frac{b + 1}{b_d + 1}}, \\
    \e_v &= (\frac{1}{T_v} - 1)T(b - 1),
\end{align}
then it becomes clear that the total effect of the noisy communications channel, postprocessing and rescaling in the `real' protocol is equivalent to the effective channel 
\begin{align}
    \mathcal{E}_T(T_\text{tot}, \e_\text{tot}) &\equiv \mathcal{E}(T, \e) \otimes \mathcal{E}_v(T_v, \e_v) \\
    &= \mathcal{E}(T T_v, \e + \frac{\e_v}{T}),
\end{align} composed of the original physical channel $\mathcal{E}$ and a virtual channel $\mathcal{\hat P} = \mathcal{E}_v$ representing Bob's virtual operations.

We note here that restricting the renormalisation to not illegally increase correlations in \eqref{eq:bBenchmark_physicalityinequality} translates straightforwardly into a restriction on the virtual channel $\mathcal{E}_v$ being physical, i.e. having $T_v \leq 1$. Again, since the physicality of the equivalent channel $\mathcal{E}_\text{tot}$ is now guaranteed, Alice and Bob may legitimately distil a shared secret-key from their postprocessed and renormalised data according to the covariance matrix \eqref{eq:bBenchmarkCV}. We note also that choosing $\Delta_V$ such that $b' = b$ allows for a higher secret-key fraction than choosing $\Delta_V$ such that $c' = c$, as described in Section \ref{sec:results}, but we stress that our choice of $\Delta_V$ is not guaranteed to be optimal. We therefore suggest that more efficient choices of $\Delta_V$ that also satisfy the requirement of physicalising the postprocessed data could exist, but we do not discuss them here.

\section{Asymptotic key rate calculation} \label{app:asym_keyrate}
Under reverse reconciliation, the minimum guaranteed secret-key bits per pulse $K$ in the asymptotic regime is given most generally by \cite{laudenbach_continuousvariable_2018}
\begin{align}
    K &= \beta I_{AB} - \chi_{EB},
\end{align}
for information reconciliation efficiency $\beta$, continuous mutual information between Alice and Bob $I_{AB}$, and Holevo information between Eve and Bob $\chi_{EB}$. 

However, because the joint measurement distribution obtained by Alice and Bob is non-Gaussian, it is difficult to calculate $K$ directly \cite{leverrier_continuous-variable_2011}. Instead, we exploit the optimality of Gaussian attacks \cite{navascues_optimality_2006, garcia-patron_unconditional_2006} to place an upper bound on the Holevo information achievable by Eve. Assuming that Eve holds a purification $\hat \rho^{ABE}$ of the state $\hat \rho_G^{AB}$, where $\hat \rho_G^{AB}$ is the Gaussian state with first- and second-order moments equal to Eqs. \eqref{eq:mABdispP} and \eqref{eq:cvABdispP} respectively, then we can compute
\begin{align}
    \chi_{EB} &\leq \chi_{EB} ({\hat \rho^{ABE}}) \\
    &= S(\hat \rho^E) - S(\hat \rho^{E|B}) \\
    &= S(\hat \rho_G^{AB}) - S(\hat \rho_G^{A|B}).
\end{align}
This represents the maximum information Eve can obtain from the protocol under collective attacks \cite{garcia-patron_unconditional_2006}. If we consider an arbitrary non-Gaussian protocol characterised to Gaussian order by the covariance matrix
\begin{align}
    \bs{V}_{AB} &= \begin{pmatrix}
        a\id & c\sz \\
        c\sz & b\id
    \end{pmatrix},
\end{align}
then the von Neumann entropies $S(\hat \rho_G^{AB})$ and $S(\hat \rho_G^{A|B})$ may be calculated from the symplectic eigenvalues $\lambda_{1,2}$ and $\lambda_3$ of the covariance matrices $\bs{V}_{AB}$ and $\bs{V}_{A|B}$ respectively \cite{laudenbach_continuousvariable_2018}:
\begin{align}
    S(\hat \rho_G^{AB}) &= G(\lambda_1) + G(\lambda_2), \\
    S(\hat \rho_G^{A|B}) &= G(\lambda_3),
\end{align}
where
\begin{align}
    \lambda_{1,2} &= \sqrt{\frac{1}{2} \LL( D_1 \pm \sqrt{D_1^2 - 4 D_2^2} \RR)}, \\
    D_1 &= a^2 + b^2 - 2c^2, \\
    D_2 &= a b - c^2
\end{align}
and
\begin{align}
    \lambda_3 &= a - \frac{c^2}{b + 1}
\end{align}
for the $G$-function
\begin{align}
    G(x) &= \frac{x + 1}{2}\log_2(\frac{x + 1}{2}) - \frac{x - 1}{2}\log_2(\frac{x - 1}{2}).
\end{align}
The extremality properties of Gaussian states \cite{wolf_extremality_2006} also allows the mutual information $I_{AB}$ to be bounded below by the information associated with the equivalent Gaussian-moment state $\hat \rho_G^{AB}$:
\begin{align}
    I_{AB} &\geq I({\hat \rho_G^{AB}}) \\
    &= \log_2 \LL( \frac{a + 1}{a + 1 - \frac{c^2}{b + 1}} \RR).
\end{align}
Hence we compute a lower bound on $K$
\begin{align}
    K &\geq K^{\infty} \\
    &= \beta I({\hat \rho_G^{AB}}) - \chi_{EB} ({\hat \rho^{ABE}}).
\end{align}

\section{Parameter estimation and security  parameters in the finite-size protocol} \label{app:fnsz_keyrate}
For an arbitrary Gaussian protocol described by general covariance matrix
\begin{align}
    \bs{V}_{AB} &= \begin{pmatrix}
        a\id & c\sz \\
        c\sz & b\id
    \end{pmatrix},
\end{align}
Alice and Bob estimate their mutual information $I(\hat \rho_{G}^{AB})$ from the estimated mean values $\hat a, \hat b, \hat c$ of the covariance matrix elements
\begin{align}
    \overline{I}(\hat \rho_{G}^{AB}) &= I(\hat \rho_{G}^{AB})\bigg|_{\LL\{a = \hat a, \ b = \hat b, \ c = \hat c\RR\}}.
\end{align}
In order to avoid underestimating Eve's information, the Holevo quantity is then estimated using the worst-case confidence interval estimates $\LL\{\Sigma_a^{min}, \Sigma_b^{min}, \Sigma_c^{max}\RR\}$ of the covariance matrix elements, such that the true Holevo quantity only exceeds the estimated value with probability $\epsilon_\text{PE}$:
\begin{align}
    f(&\Sigma_a^{max}, \Sigma_b^{max}, \Sigma_c^{min}) \notag\\
    &= \chi_{EB} ({\hat \rho^{ABE}})\bigg|_{\LL\{a = \Sigma_a^{max}, \ b = \Sigma_b^{max}, \ c = \Sigma_c^{min}\RR\}}.
\end{align}
The worst-case estimates are given by 
\begin{align}
    \Sigma_a^{max} &= \LL (1 + \delta_\text{Var} \RR) \hat a, \\
    \Sigma_b^{max} &= \LL (1 + \delta_\text{Var} \RR) \hat b, \\
    \Sigma_c^{min} &= \LL (1 - 2 \sqrt{\frac{\hat a \hat b}{\hat c^2}} \delta_\text{Cov} \RR) \hat c,
\end{align}
where
\begin{align}
    \delta_\text{Var} &= \LL[2 - A\LL( \frac{\epsilon_\text{PE}}{12} \RR) \RR] \LL[ 1 + \frac{240}{\epsilon_\text{PE}}e^{-N/32} \RR] - 1, \\
    \delta_\text{Cov} &= \frac{1}{2}\LL[ 1 - A\LL( \frac{\epsilon_\text{PE}}{12} \RR) \RR] + \LL[ 1 - A\LL( \frac{\epsilon_\text{PE}^2}{1296} \RR) \RR]
\end{align}
and
\begin{align}
    A(z) &= 2 \ \text{invcdf}_{\text{Beta} \LL[\frac{N}{2}, \frac{N}{2} \RR]} \LL( z \RR).
\end{align}
For ease, we choose our security parameters, and other relevant parameters in our protocol, largely in accordance with Ref. \cite{jain_practical_2022}. These are summarised in Table \ref{tab:security_params}.

\begin{table*}[!htb]
\centering
\renewcommand{\arraystretch}{1.5}
\begin{tabular}{c|l|c}
\textbf{Parameter} & \textbf{Description}                                   & \textbf{Value} \\ \hline
$\beta$                     & Gaussian data reconciliation efficiency                                                     & 0.95                          \\
$d_{rx}$                    & Number of discretization bits for ADC of Gaussian data                                      & 6                             \\
$p_f$                       & Success probability of error correction per frame                                           & 0.9964                        \\
$\epsilon_\text{ent}$            & Confidence level in estimating the entropy $H(\overline{Y})$ of the discretized quantum symbols $\overline{Y}$ & $10^{-10}$ \\
$\epsilon_s$                & Min-entropy smoothing parameter                                                         &  $10^{-10}$                   \\
$\epsilon_h$                & Leftover hash lemma confidence parameter                       & {$10^{-10}$}                  \\
$\epsilon_\text{PE}$             & Confidence level of parameter estimation                                                    & {$10^{-10}$}   
\end{tabular}
\caption{A list and brief description of the security parameters used to estimate the finite-size secret key fractions $K^\mathcal{F}$ in Section \ref{sec:results_finitesize}. We note that some $\epsilon$ parameters, e.g. $\epsilon_{qrng}$, are not listed. These parameters are contained implicitly within other factors in the protocol and do not need to be explicitly specified \cite{jain_practical_2022}.}
\label{tab:security_params}
\end{table*}

\section{Satellite-based platforms}

The use of satellite-based platforms to extend the viable distance of quantum communications has been widely regarded as one of the most promising solutions towards a global-scale quantum Internet. To this end, SQCC protocols may find advantage on satellite-based platforms, where size and power usage are restricted. The use of satellite-based free-space-optical channels also eliminates the issue of fibre attenuation on the viable distance of CV-QKD, rendering SQCC an attractive solution to be incorporated into standard satellite-based optical classical communications systems.  

Satellite-based implementation of SQCC faces several distinct challenges. Of these, atmospheric-turbulence-induced fluctuations of channel transmissivity resulting from the optical propagation within the Earth's turbulent atmosphere~\cite{andrews_laser_2005} represent one of the most critical obstacles. Due to the nature of joint quantum-classical modulation, both Alice and Bob require precise information on the instantaneous channel transmissivity. Indeed, Alice requires this information in order to jointly optimise the large phase-space displacement $\tilde{d}$ and the QKD modulation variance $V$ for her encoding, such that the secret key rate is maximised while the classical quality-of-service is met. Bob also requires this information to perform the re-displacement and determine the electronic scaling gain for his decoding. Without such information, Alice and Bob may resort to using the statistical average of the channel transmissivity for their purposes; however, this will greatly reduce the feasibility of SQCC. Even in the typical satellite-to-Earth (i.e., downlink) configuration where the channel fluctuations can be considered minimal (see, e.g.,~\cite{winnel_classical-quantum_2024}), failing to precisely erase the effect of the large displacement and emulate the measurement outcomes that would have been generated had Alice performed ordinary zero-mean GMCS CV-QKD at Bob's end will introduce a unique fluctuation-induced QKD decoding noise whose contribution is proportional to the classical power. Our detailed simulations of satellite-based atmospheric optical propagation reveal that the contribution of this QKD decoding noise can be much more significant than that of the phase noise even under the minimal channel fluctuations as encountered in a typical downlink channel, effectively rendering space-based SQCC \textit{based on channel statistics} non-viable  (results are not presented here).  

The aforementioned negative outcome can be reversed if information on the instantaneous channel transmissivity is available to the system. The requirement of precise information on the instantaneous channel transmissivity necessitates real-time channel probing for the successful implementation of SQCC in a real-world satellite-based deployment. In the most intuitive implementation, this information is acquired by Bob via his measurement of the local oscillator, which is typically sent by Alice along with the combined quantum-classical signal for the establishment of a phase reference. Bob can then share this information with Alice via some feedback mechanism. 

However, the update of channel information is limited by the satellite-to-Earth round-trip time, which is ${\sim} 3.3\,\text{ms}$, assuming that a typical low-Earth-orbit satellite (positioned at an altitude of $500\,\text{km}$) is located at the zenith point relative to the ground station. 
Since the coherence time of an atmospheric channel is typically on the order of $1\,\text{ms}$~\cite{andrews_laser_2005}, such an implementation can be very challenging. 
Nevertheless, while the requirement of precise information on the instantaneous channel transmissivity is strictly necessary at Bob, it is expected that this requirement can be relaxed at Alice in a downlink scenario (where channel fluctuations are minimal). Therefore, SQCC should be suitable for the downlink configuration, which represents the mainstream configuration adopted in satellite-based optical classical and quantum communications. It should also be noted that SQCC is naturally well-suited for the inter-satellite configuration (an important yet less explored scenario for communications within a satellite constellation) due to the absence of atmospheric turbulence in space. To summarize, within the context of satellite-based communications, the results presented in this work will be useful for both satellite-to-Earth (downlink) and inter-satellite configurations. They will not be useful for the Earth-to-satellite (uplink) configuration. 

In addition, this work does not consider the use of classical channel coding, which effectively drives the classical bit error rates to zero and eliminates one aspect of the classical-quantum interplay. If an appropriate classical channel coding scheme is used, the impact of the non-Gaussian noise contribution related to the classical bit error will be significantly reduced. Nonetheless, investigation of SQCC without channel coding is still useful as it allows for new insights into the role played by the classical bit error rate on the allowed secret key rates and allows for more direct comparisons with the existing literature.

\bibliography{references.bib}

\end{document}